\documentclass{aa}

\usepackage{natbib}
\usepackage{amsmath}
\bibpunct{(}{)}{;}{a}{}{,} 
\usepackage{graphics}   
\usepackage{graphicx}   
\usepackage{epstopdf}
\usepackage{longtable}   
\usepackage{url}      
\usepackage{bm}        
\usepackage[varg]{txfonts}
\usepackage{pdflscape}
\usepackage{supertabular}
\usepackage{hyperref}   
\usepackage[svgnames]{xcolor}
\usepackage{longtable}
\usepackage{lscape}
\usepackage{booktabs}
\usepackage{graphicx}
\usepackage{multicol}
\usepackage{placeins}
\usepackage{afterpage}
\newcommand{\Msol}{M\ensuremath{_{\sun}}}

\usepackage{float}

\begin{document}

\title{Very Massive Stars Models} 
\subtitle{I. Impact of Rotation and Metallicity and Comparisons with
Observations}

\author{S\'ebastien Martinet\inst{1,2}, Georges Meynet\inst{1}, Sylvia Ekstr{\"o}m\inst{1}, Cyril Georgy\inst{1}, Raphael Hirschi\inst{3,4}}

 \institute{ Geneva Observatory, University of Geneva, Chemin Pegasi 51, CH-1290 Versoix, Switzerland
 \and Institut d’Astronomie et d’Astrophysique, Universit\'e Libre de Bruxelles, CP 226, B-1050 Brussels, Belgium\\
 email: sebastien.martinet@ulb.be 
 \and Astrophysics Group, Keele University, Keele, Staffordshire, ST5 5BG, UK
 \and Institute for Physics and Mathematics of the Universe (WPI), University of Tokyo, 5-1-5 Kashiwanoha, Kashiwa 277-8583, Japan
}
 \authorrunning{Martinet et al.}

\date{Accepted August 31, 2023}

\abstract{In addition to being spectacular objects, Very Massive Stars (VMS) are suspected to have a tremendous impact on their environment and on the whole cosmic evolution. The nucleosynthesis both during their advanced stages and their final explosion may contribute greatly to the overall enrichment of the Universe. Their resulting supernovae are candidates for the most superluminous events and their extreme conditions also lead to very important radiative and mechanical feedback effects, from local to cosmic scale. }
{We explore the impact of rotation and metallicity on the evolution of very massive stars across cosmic times.}
{With the recent implementation of an equation of state in the GENEC stellar evolution code, appropriate for describing the conditions in the central regions of very massive stars in the advanced phases, we present new results on VMS evolution from Population III to solar metallicity. }
{Low metallicity VMS models are highly sensitive to rotation, while the evolution of higher metallicity models is dominated by mass loss effects. The mass loss affects strongly their surface velocity evolution, breaking quickly at high metallicity while reaching the critical velocity for low metallicity models. The comparison to observed VMS in the LMC shows that the mass loss prescriptions used for these models are compatible with observed mass loss rates. In our framework for modelling rotation, our models of VMS need a high initial velocity to reproduce the observed surface velocities. The surface enrichment of these VMS is difficult to explain with only one initial composition, and could suggest multiple populations in the R136 cluster. At a metallicity typical of R136, only our non- or slowly rotating VMS models may produce Pair Instability supernovae. The most massive black holes that can be formed are less massive than about 60 M$_\odot$.
}
{Direct observational constraints on VMS are still scarce. Future observational campaigns will hopefully gather more pieces of information guiding the theroretical modeling of these objects whose impacts can be very important. VMS tables are available online.}


\maketitle

\section{Introduction}
Very massive stars (VMS), often considered as stars with initial mass larger than 100 \Msol, have been mostly stimulating interest in Population III and early Universe. Indeed, early hydrodynamical simulations predicted preferential formation of very massive stars for Population III stars \citep{Abel2002,Bromm2002}. More recent simulations predict now a more significant fragmentation process \citep{Stacy2010,Clark2011}, binaries fraction \citep{Turk2009} and wide initial mass distributions \citep{Hirano2014,Hirano2015}. Moreover, the IMF might even be redshift-dependent for Population III stars \citep{Hirano2015} due to the different temperature of formation sites. There is, however, a general agreement on a top-heavy primordial IMF \citep{Greif2011,Stacy2013,Hirano2014,Hirano2015,Susa2014,Stacy2016,Jerabkova2018,Wollenberg2020}. Population III stars could be a very significant source of ionizing photons and thus be interesting candidates for the reionization of the intergalactic medium \citep{Haehnelt2001,FaucherGiguere2008,FaucherGiguere2009, Becker2013,Wise2014,Sibony2022}. Using the present Pop III models, \citet{Murphy2021} showed that including initial masses up to 500 M$_\odot$ can increase the total number of ionizing by 30\% compared to the case where the upper mass limit is chosen equal to 120 M$_\odot$.
While VMS have an undeniable impact in the early universe, their contribution across the cosmic time might be underestimated even in higher metallicity environments. 

One of the most spectacular observations in the early 80s was the observation in the 30 Doradus Nebulae, in the Large Magellanic Cloud (LMC), of the central object R136, that \citet{Cassinelli1981} showed to potentially be a $\sim$2500 \Msol\ star according to their spectroscopic analysis. With higher spatial resolution instruments, R136 was then shown later on to be a young cluster, composed of much lower mass stars. The stellar upper mass limit was then found to lie around 150 \Msol\ 
\citep{Weidner2004,Figer2005,Oey2005,Koen2006}. 
\cite{Crowther2010} used new adaptive optics observations combined with more refined non-LTE atmosphere code to derive the fundamental parameters of these objects. Their results show that for four stars in R136, models suggest masses in the 165-320\Msol\ mass range. This has been confirmed by recent studies using new HST data \citep{Bestenlehner2020,Brands2022}. 

Interestingly, VMS could also be viable candidates to explain the self-enrichment of globular clusters \citep[see ][]{Vink2018}. For the last decade, one of the main candidates for this self-enrichment process has been massive AGB stars \citep{DErcole2010} due to their slow winds compare to the fast line-driven winds of massive O-stars. Indeed, such fast winds (up to 2000-3000 km.s$^{-1}$) are not able to deposit the enriched material efficiently into the globular clusters as the winds are too fast to be trapped in their potential well. Alternatives such as rotating massive stars \citep{Decressin2007}, massive binaries {\citep{deMink2009,Vanbeveren2012}}, red supergiants \citep{Szecsi2018}, or supermassive (SMS) stars \citep{Denis2014} have been proposed and are discussed in \citet{Bastian2018}. Interestingly, SMS have been proposed as a workaround to the so-called mass-budget problem \citep{Gieles2018}, but the estimated wind velocities of roughly 1000 km s$^{-1}$ are still too fast compared to the estimated escape velocities ($\simeq$ 500 km s$^{-1}$) at the center of young globular clusters \citep[][]{Gieles2018}. \citet{Vink2018} propose then that VMS, after inflating due to their proximity to the Eddington limit \citep{Ishii1999,Graefener2012,Sanyal2015}, lose mass through slower winds, satisfying conditions on both wind velocity for enrichment and mass-budget.

Moreover, VMS have been shown to have a potential important impact for the chemical enrichment at high metallicity. Indeed, \citet{Martinet2022} have shown that the galactic production of the short-lived radioactive isotope $^{26}$Al by the wind-driven mass loss of massive stars might be increased by up to 150\% when including VMS into a Milky Way-like population, and could then account for a large part of the observed $^{26}$Al galactic content.

To explore the evolution of VMS across the cosmic history, and as a follow-up to the work done in \citet{Yusof2013}, we computed a grid of both rotating and non-rotating VMS from Population III stars to solar metallicity stars with GENEC \citep{Eggenberger2008}. In this work, we take advantage of the improvements made compared to \citet{Yusof2013} models, such as the improvement of angular momentum conservation \citep{Ekstrom2012}, the use of another convective criterion \citep{Kaiser2020}, a revised overshooting parameter \citep{Martinet2021} and the introduction in this work of an equation of state accounting for the pair-creation production.

 The paper is organized as follows: in Section 2 we describe the physical ingredients used to compute the present models for very massive stars. The effect of different physics on very massive stars evolution is discussed in Section 3. 
Comparison between our models and observed VMS in the LMC is discussed in Section 4. In Section 5 we discuss and suggest potential impact of different physics on the final fate of VMS, and we present the main conclusions of this work.




\section{Ingredients of the stellar models}

The present models have been computed with the 1D stellar evolution code GENEC \citep{Eggenberger2008}. These models
differ mainly in three points with respect to the physics used in the grids by \cite{Ekstrom2012} and \cite{Yusof2022}. First, we adopted the Ledoux criterion for convection
instead of Schwarzschild and an overshoot of 0.2 $H_p$ instead of 0.1 $H_p$. These changes were made because there are some indications that the Ledoux criterion might be more appropriate for reproducing some observed properties of massive stars \citep[][]{Georgy2014, Kaiser2020}, and that an increase in overshoot parameter is needed for stars with masses higher than 8 \Msol\ \citep{Martinet2021,Scott2021}. The third change is that our very massive star models have been computed with an equation of state accounting for electron-positron pair production \citep{Timmes2000}. 
The other ingredients are the same as in \cite{Ekstrom2012} and \cite{Yusof2022}. 
To make the paper more self-consistent, let us however remind the prescriptions used for rotation and mass loss.
We considered here the physics of shellular rotating models as described in \citet{Zahn1992}. For the vertical shear diffusion coefficient, we used the expression by \citet{Maeder1997} and the expression of the horizontal shear diffusion coefficient by \citet{Zahn1992}. 


\label{Sect:Ingredients_of_models}

The radiative mass-loss rate adopted on the MS is from \citet{Vink2001}; the domains not covered by this
prescription { \citep[see Fig. 1 of][]{Eggenberger2021}} use the \citet{deJager1988} rates. 
\citet{Grafener2008} prescriptions are used in their domain of application, while \citet{Nugis2000} prescriptions are used everywhere else for the Wolf-Rayet phase\footnote{The Wolf-Rayet phase in GENEC is assumed to begin when the model has an effective temperature larger than 10 000 K and a surface mass fraction of hydrogen below 0.3.}. We kept here for consistency the same prescriptions as for the overall grids published so far, however more recent prescriptions are now available \citep[see the review of][and Fig. \ref{fig:Mdot_Gamma_Edd_Brands2022}]{Vink2021}. {In Sect. \ref{sect:conclusion_physical_ingredients}, we propose a more in-depth discussion on new mass loss prescriptions and the relevance of the prescriptions used in this work. }
The radiative mass-loss
rate correction factor described in \citet{Maeder2000} is applied for rotating models. 
The dependence on metallicity is taken such that  $\dot{M}$(Z) = (Z/Z$_\odot$)$^{0.7} \dot{M}$(Z), except during
the red supergiant (RSG) phase, for which no dependence on the metallicity is used. This is done accordingly to \citet{vanLoon2005} and \citet{Groenewegen2012a,Groenewegen2012b} showing that the metallicity dependence for the mass loss rates of these stars do appear to be weak.

We computed stellar models with initial masses of 180, 250, and 300 \Msol\ for Z=0.000, Z=10$^{-5}$, Z=0.006 and Z=0.014 metallicities, with no rotation and a rotation rate of V/V$_c$=0.4, where V$_c$ is the critical velocity\footnote{The critical velocity is the velocity at which the centrifugal force at the equator balances the gravity there. Its expression is taken as indicated by expression 6 in \citet{Eks2008}.}. The nuclear network allows following the abundance variation of 30 isotopes \footnote{These isotopes are  $^{1}$H, $^{3,4}$He, $^{12,13,14}$C, $^{14,15}$N, $^{16,17,18}$O, $^{18,19}$F, $^{20,21,22}$Ne, $^{23}$Na, $^{24,25,26}$Mg, $^{26,27}$Al, $^{28}$Si, $^{32}$S, $^{36}$Ar, $^{40}$Ca, $^{44}$Ti, $^{48}$Cr, $^{52}$Fe, and $^{56}$Ni.}. 

\section{The grid of VMS stellar models}

    \begin{table*}
    \centering
    \caption{Properties of the stellar models at different stages of their evolution. The first columns present the initial mass M$_{\rm ini}$ in \Msol, the initial surface velocity $\nu_{\rm ini}$ in km.s$^{-1}$ and the averaged surface velocity over the main-sequence $\Bar{\nu}_{\rm MS}$ in km.s$^{-1}$. For the H-, He- and C- burning phases, the columns present the durations of the phase $\tau_{\rm phase}$ in Myrs for H- and He- and in yrs for C-, the current total mass M$_{\rm tot}$ in \Msol, the surface velocity $\nu_{\rm surf}$ in km.s$^{-1}$ and the surface mass fraction of helium Y$_{\rm surf}$ at the end of each nuclear burning phases. }
    \begin{tabular}{p{0.72cm}p{0.72cm}p{0.72cm}p{0.72cm}p{0.72cm}p{0.72cm}p{0.72cm}p{0.72cm}p{0.72cm}p{0.72cm}p{0.72cm}p{0.72cm}p{0.72cm}p{0.72cm}p{0.72cm}p{0.72cm}}
\toprule
\toprule
\multicolumn{3}{c|}{\textbf{Initial parameters}} & \multicolumn{4}{c|}{\textbf{End of H-burning}} & \multicolumn{4}{c|}{\textbf{End of He-burning}}     & \multicolumn{4}{c}{\textbf{End of C-burning}}       \\
  M$_{\rm ini}$ &  $\nu_{\rm ini}$ &  $\Bar{\nu}_{\rm MS}$ &  $\tau_{\rm H}$ & M$_{\rm tot}$ &  $\nu_{\rm surf}$ &  Y$_{\rm surf}$ & $\tau_{\rm He}$ &     M$_{\rm tot}$ &  $\nu_{\rm surf}$ &  Y$_{\rm surf}$ &  $\tau_{\rm C}$ & M$_{\rm tot}$ &  $\nu_{\rm surf}$ &  Y$_{\rm surf}$ \\

\midrule
\multicolumn{15}{c}{Z=0.014}                                                                                                                                                                                  \\ 
\midrule
180         & 0.0           & 0.0                & 2.402  & 92  & 0.0            & 0.89           & 0.283         & 48  & 0.0            & 0.21         & 2.752        & 48  & 0.0            & 0.20          \\
250         & 0.0           & 0.0                & 2.236  & 99  & 0.0            & 0.99           & 0.289         & 44  & 0.0            & 0.22         & 3.997        & 44  & 0.0            & 0.22          \\
300         & 0.0           & 0.0                & 2.160  & 81  & 0.0            & 0.98           & 0.296         & 39  & 0.0            & 0.24         & 3.453        & 39  & 0.0            & 0.23          \\
180         & 430           & 110                & 2.622  & 87  & 1.2            & 0.92           & 0.285         & 43  & 5.3            & 0.24         & 2.688        & 42  & 24.5           & 0.23          \\
250         & 446           & 87                 & 2.439  & 69  & 1.2            & 0.98           & 0.310         & 28  & 1.7            & 0.25         & 5.975        & 28  & 4.7            & 0.23          \\
300         & 455           & 78                 & 2.269  & 77  & 0.7            & 0.98           & 0.292         & 37  & 1.3            & 0.23         & 4.708        & 37  & 5.0            & 0.22      \vspace{0.1cm}    \\ 
\midrule
\multicolumn{15}{c}{Z=0.006}                                                                                                                                                                                  \\ 
\midrule
180         & 0.0           & 0.0                & 2.384  & 114 & 0.0            & 0.77           & 0.267         & 71  & 0.0            & 0.25         & 1.408        & 71  & 0.0            & 0.24          \\
250         & 0.0           & 0.0                & 2.190  & 154 & 0.0            & 0.84           & 0.250         & 109 & 0.0            & 0.23         & 0.493        & 109 & 0.0            & 0.22          \\
300         & 0.0           & 0.0                & 2.104  & 180 & 0.0            & 0.89           & 0.269         & 91  & 0.0            & 0.23         & 0.553        & 91  & 0.0            & 0.22          \\
180         & 454           & 221                & 2.631  & 75  & 2.4            & 0.99           & 0.282         & 46  & 4.6            & 0.28         & 2.582        & 45  & 26.3           & 0.27          \\
250         & 481           & 206                & 2.376  & 88  & 0.9            & 0.99           & 0.272         & 56  & 1.8            & 0.29         & 1.430        & 56  & 38.3           & 0.28          \\
300         & 492           & 197                & 2.263  & 96  & 0.6            & 0.99           & 0.27          & 60  & 1.0            & 0.27         & 1.257        & 60  & 9.0            & 0.25       \vspace{0.1cm}    \\ 
\midrule
\multicolumn{15}{c}{Z=10$^{-5}$}                                                                                                                                                                              \\ 
\midrule
180         & 0.0           & 0.0                & 2.292  & 180 & 0.0            & 0.25           & 0.24          & 111 & 0.0            & 0.85         & 1.661        & 111 & 0.0            & 0.85          \\
250         & 0.0           & 0.0                & 2.115  & 202 & 0.0            & 0.49           & 0.25          & 153 & 0.0            & 0.88         & 0.356        & 153 & 0.0            & 0.88          \\
300         & 0.0           & 0.0                & 2.036  & 219 & 0.0            & 0.67           & –             & –   & –              & –            & –            & –   & –              & –             \\
180         & 574           & 572                & 2.740  & 174 & 101.0          & 0.71           & 0.24          & 173 & 6.4            & 0.83         & –            & –   & –              & –             \\
250         & 617           & 613                & 2.443  & 240 & 112.0          & 0.83           & 0.23          & 237 & 5.5            & 0.90         & 0.355        & 237 & 6.3            & 0.90          \\
300         & 648           & 648                & 2.330  & 287 & 356.0          & 0.83           & 0.22          & 274 & 0.6            & 0.90         & –            & –   & –              & –          \vspace{0.1cm}    \\ 
\midrule
\multicolumn{15}{c}{Z=0.000}                                                                                                                                                                                  \\ 
\midrule
180         & 0.0           & 0.0                & 2.875  & 180 & 0.0            & 0.25           & –             & –   & –              & –            & –            & –   & –              & –             \\
250         & 0.0           & 0.0                & 3.083  & 250 & 0.0            & 0.25           & –             & –   & –              & –            & –            & –   & –              & –             \\
300         & 0.0           & 0.0                & 3.012  & 300 & 0.0            & 0.25           & –             & –   & –              & –            & –            & –   & –              & –             \\
180         & 768           & 782                & 2.323  & 176 & 587.0          & 0.43           & –             & –   & –              & –            & –            & –   & –              & –         \vspace{0.1cm}     \\
\midrule
\end{tabular}
    \label{table:models_details-1}
    \end{table*}

Table \ref{table:models_details-1} presents the initial parameters and various physical quantities of the models presented in this paper at different {evolutionary} stages. We define the beginning of a phase when 1\% of the central burning element has been consumed and the end of a phase {when the 
mass fraction of the main fuel at the center} is lower than 10$^{-3}$. 

\begin{figure*}
    \centering
    \includegraphics[width=0.9\textwidth]{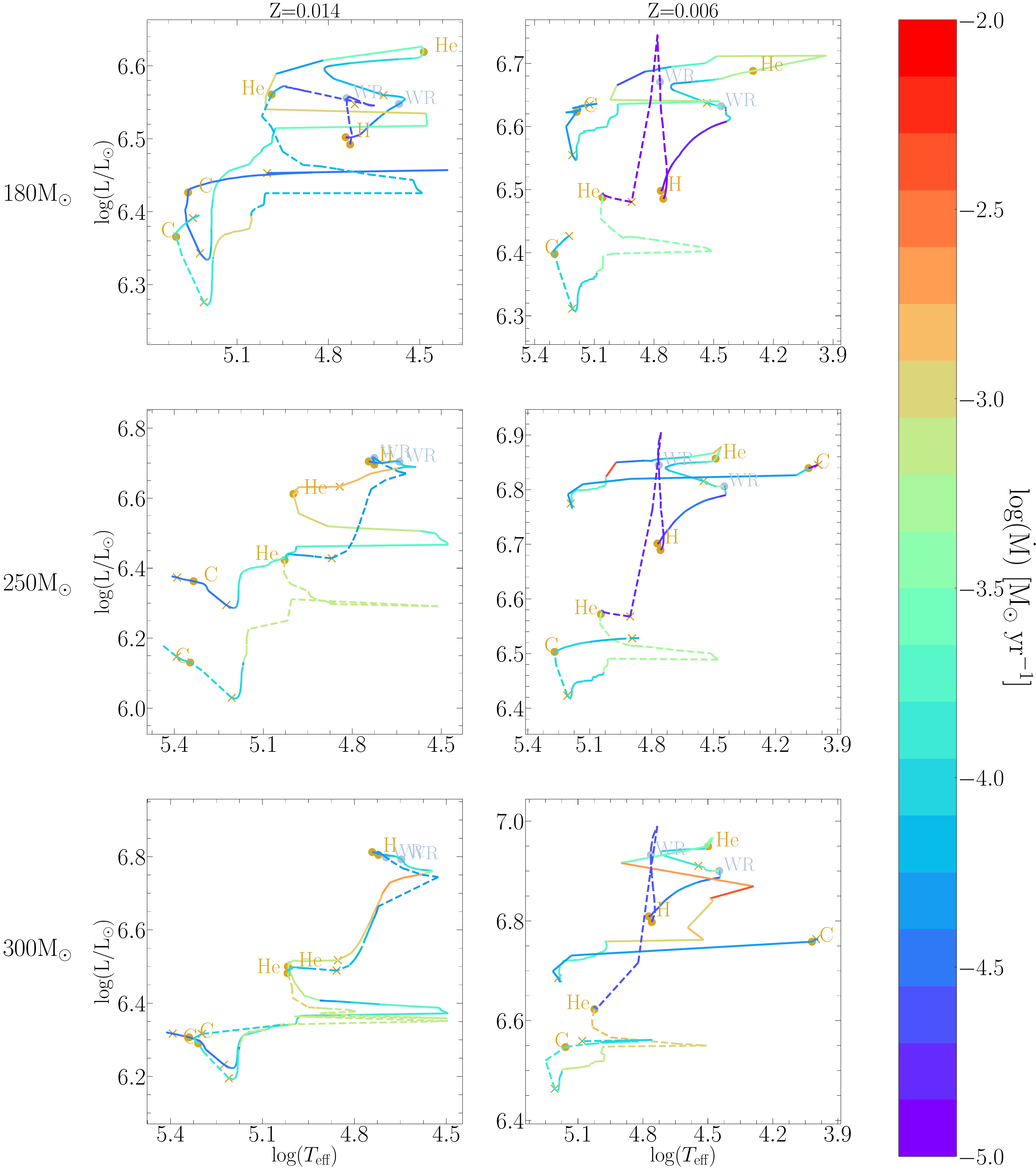}
    \caption{HRD of the non-rotating (solid lines) and rotating at V/V$_c$=0.4 (dashed lines) models at Z=0.014 (solar) and Z=0.006 (LMC). The top/middle/bottom row shows model with initial masses of 180/250/300\Msol. The tracks are color-coded according to their mass loss rates. The beginning and the end of the burning phases are indicated respectively by circles and crosses. The beginning of the Wolf-Rayet phase is indicated by light blue circles. 
    }
    \label{fig:HRD_grid_z14_z06}
\end{figure*}

Figure \ref{fig:HRD_grid_z14_z06} presents the Hertzsprung-Russell diagram (HRD) of the higher metallicities models, at Z=0.014 and Z=0.006. The non-rotating models are displayed in solid lines, while the rotating ones are in dashed lines. Starting with the 180 \Msol\ at Z=0.014, we can see that the rotating model evolves more to the blue during the MS due to the rotational mixing. {It enters 
the WR phase at a  slightly earlier  evolutionary stage. The rotating model enters the core He-burning phase at a much higher effective temperature than the non-rotating ones.} The evolution through the advanced phases is then very similar due to the dominant WR mass loss, with the rotating models at slightly lower luminosity. 

If we go to a higher initial mass, the evolution is now highly dominated by the larger mass loss, with still the rotating models evolving with slightly lower luminosities. The 300 \Msol\ at Z=0.014 is the perfect example of mass loss dominated evolution, where both the non-rotating and the rotating models undergo so much mass loss that their evolution through the HRD is effectively similar.

If we go to lower metallicity, we now have a lower contribution to the mass loss of the line-driven wind. This means that the evolution of VMS will now be more sensitive to the rotational mixing. Indeed, as we can see for the whole 180-300\Msol\ mass range, the rotating models undergo quasi-chemically homogeneous evolution due to the very strong mixing. Rotating stars reach the WR phase much earlier. This leads the star to undergo large mass loss. The He-burning phase of rotating models occurs at much lower luminosities than for the non-rotating models. The same for the more advanced phases.

To synthesize these results, we can say that at solar metallicity, the tracks of the models for the 250 and 300 M$_\odot$ nearly always decrease in luminosity as a result of the very strong mass loss. The effects of rotation become less and less strong when the initial mass increases because of the dominating effect of mass loss. At the metallicity $Z=0.006$ as well as for the 180 M$_\odot$ at solar metallicity, {\it i.e.} for all the models where the mass losses are not so strong, the effects of rotation are important. As is well known, rotating tracks are bluer on the MS phase. The models enter at an earlier evolutionary phase into the WR phase and thus reach lower luminosities during the post MS phases.

\begin{figure*}
    \centering
    \includegraphics[width=0.95\textwidth]{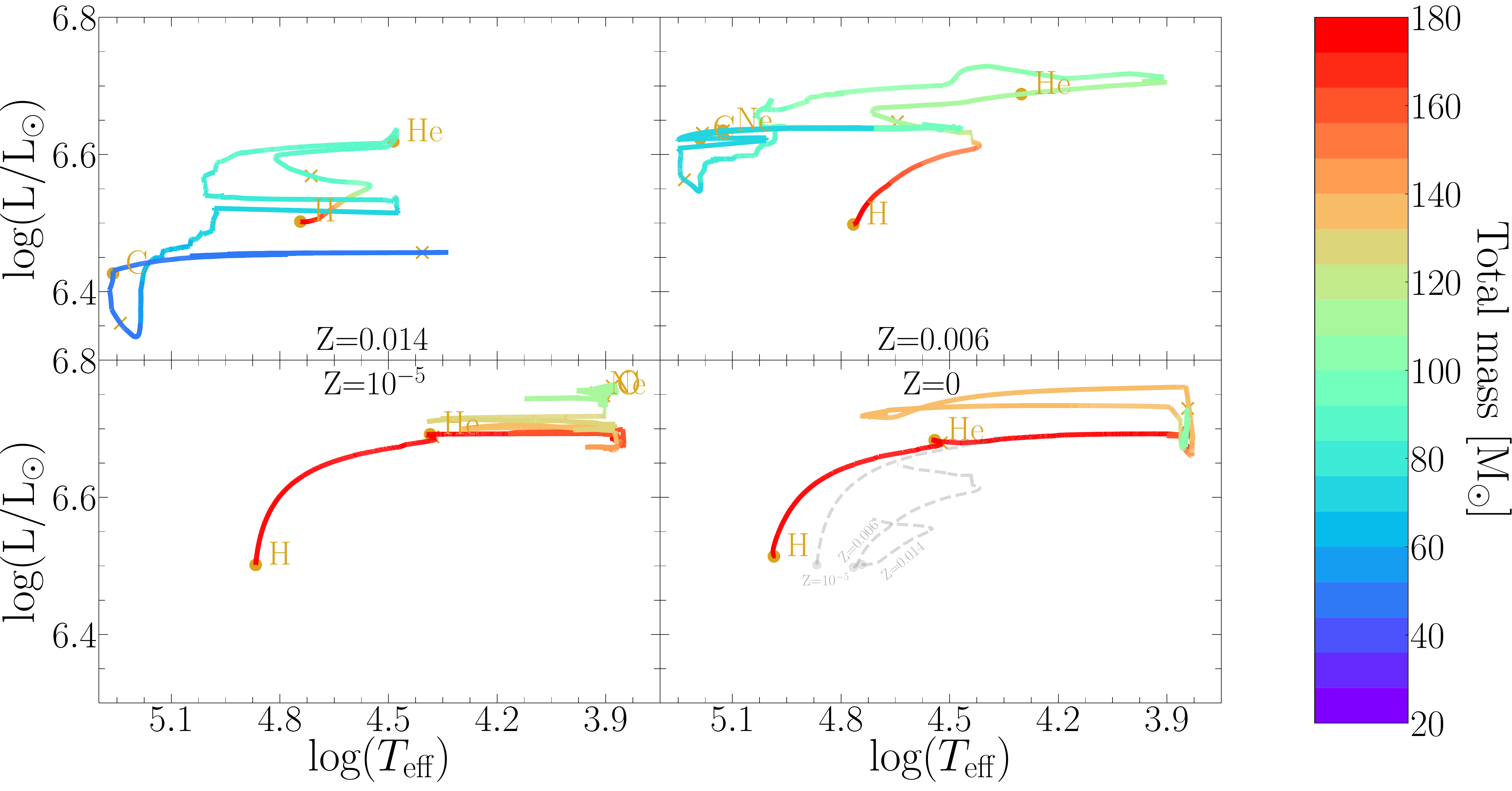}
    \caption{HRD for 180\Msol\ non-rotating models at Z=0.014, Z=0.006, Z=10$^{-5}$ and Z=0. The tracks are color-coded according to their current total mass. The MS of the models at Z=0.014, Z=0.006, Z=10$^{-5}$ are plotted in dashed grey in the lower right panel for comparison. }
    \label{fig:HRD_180Msol_z14_z06_ze5_z0}
\end{figure*}

\label{sect:impact_of_metallicity}
Figure \ref{fig:HRD_180Msol_z14_z06_ze5_z0} shows the HRD of 180 \Msol\ non-rotating stellar models at different metallicities. One of the main effects of metallicity on the surface properties is, of course, the cooler effective temperature at higher metallicity (see Fig. \ref{fig:HRD_180Msol_z14_z06_ze5_z0} bottom right panel). Indeed, we can see that the beginning of the H-burning phase, while starting at a roughly similar luminosity, starts at very different T$_{\rm eff}$. This is mainly due to the fact that the opacity of the outer layers increases with the metallicity. Since the gradient of the radiative pressure scales with the opacity, this means that the outer layers receive a stronger radiative support in high metallicity stars than in low metallicity ones. This produces stars with larger radii and lower effective temperatures.
Moreover, a higher metallicity implies more CNO elements in stars, that in turn implies that, for producing a given amount of energy, smaller temperatures are needed. Smaller temperature means that the star needs to contract less to allow nuclear reactions to produce enough energy for compensating the losses at the surface \citep[see Sect. 4.2 in][]{Farrell2022}. This also contributes to making metal-rich star more extended than metal-poor ones.

The evolution of the first parts of the MS is similar for the metallicities considered in Fig.~\ref{fig:HRD_180Msol_z14_z06_ze5_z0}, until the effect of the line-driven wind, very sensitive to metallicity, truly kicks in. From then, the higher metallicity models start to evolve to the blue due to large quantities of mass loss, quickly forming WR stars. 

At low metallicity, the star remains more massive, and this has a strong impact on the evolutionary tracks in the HRD. During the MS phase, tracks cover a larger range both in luminosity and effective temperatures. After the MS phase, since there is still an important
H-rich envelope, the star reaches larger radii. 


During the post-core He-burning phases, the low metallicity models have larger luminosities than the more metal-rich models. This is a consequence of their larger actual masses. Their effective temperature is not reaching as large values as the stripped core resulting from the higher metallicity models.
It is interesting to note that these models are very close to the Eddington limit and hence undergo large mass loss events for a short time during the advanced phases.

\begin{figure}
    \centering
    \includegraphics[width=0.49\textwidth]{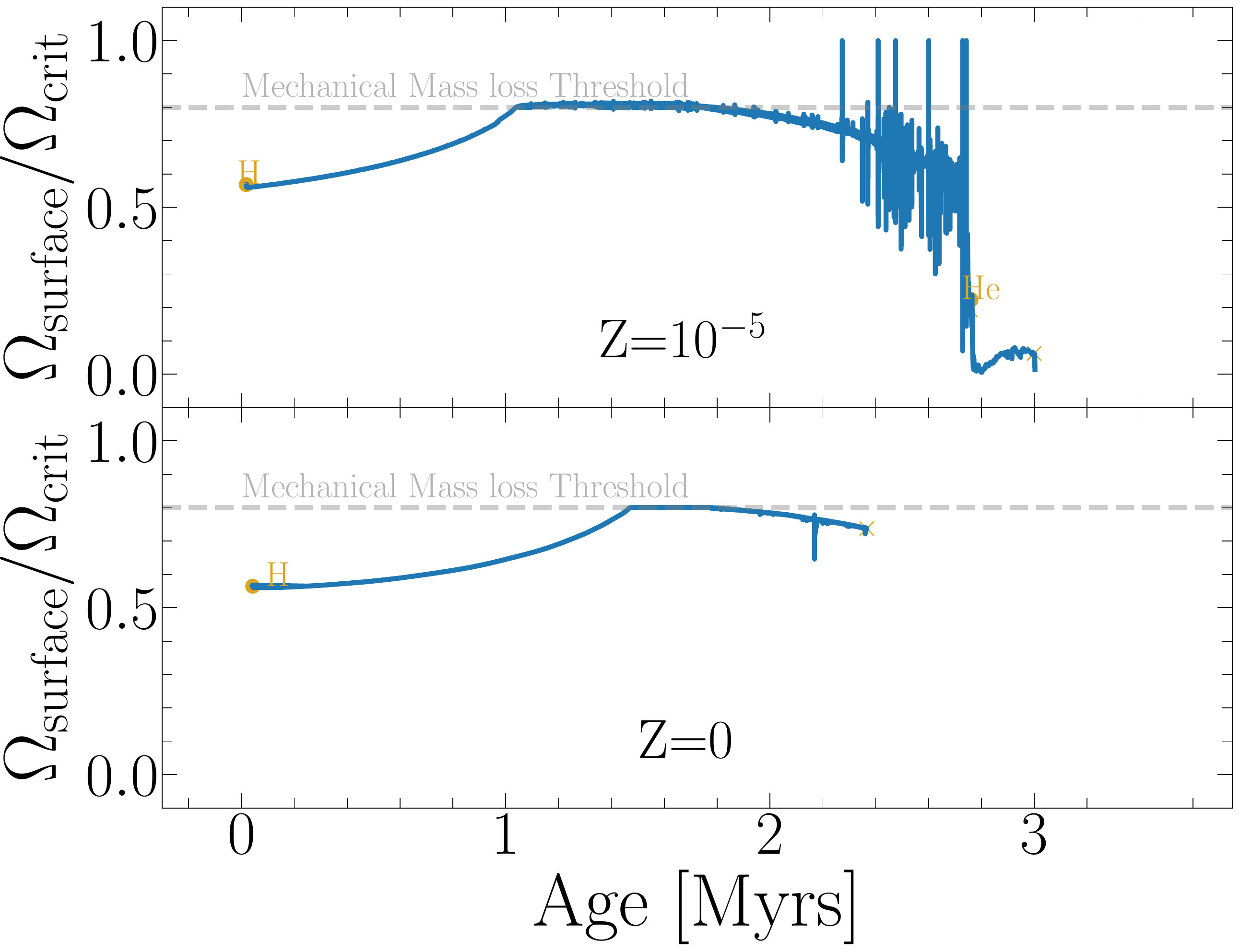}
    \caption{Evolution of the surface angular velocity divided by the critical angular velocity for 180 \Msol\ rotating models initially at V/V$_c$=0.4 ($\Omega/\Omega_{\rm crit}\simeq$0.58), for Z=10$^{-5}$ and Z=0. In grey dashed line is shown the numerical limit used to trigger the mechanical mass loss.}
    \label{fig:Vsurf_180Msol_z14_z06_ze5_z0}
\end{figure}

Figure \ref{fig:Vsurf_180Msol_z14_z06_ze5_z0} shows the evolution of the surface rotation for a 180 \Msol\ at low metallicities. 
At these metallicities, the line-driven wind is almost negligible, hence the transport of angular momentum from the core spin-up the surface. In fact, the surface accelerates up to the critical velocity. From this point on, works usually assume that some matter will be lost from the star, forming an equatorial disk \citep[see for example][]{Hirschi2007}. Indeed, at the critical velocity, a small additional force is sufficient to launch the matter in Keplerian orbits around the star. The disk may lose mass \citep[see e.g.][]{Kri2011}. The net result will be a mechanical loss of mass by the star. 

Numerically, this induces large problems of angular momentum conservation at the surface, that are very difficult to overcome when so close to the critical velocity. To manage this problem, we chose to artificially lower the effective velocity needed to trigger this mechanical mass loss to $\Omega_{\rm limit}/\Omega_{\rm crit}$=0.8. While the effects of lowering this limit are minimal in this case ($\approx$2\% of the total mass is lost by mechanical mass loss) and help tremendously the computation, it can be partly justified also by the uncertainties of the mass loss processes for very fast rotators. If we consider the case of the Be stars, which are stars showing a decretion disk likely resulting from their high rotation, it is not clear at which velocity those stars begin to lose mass \citep[see the review by][and references therein]{Rivinius2013}. Indeed, 
in addition to the centrifugal force, some other forces should be involved to kick the matter into an outward-expanding disk. It may be momentum given by radiation, or by pulsations. At least, it is likely that stars may begin to lose mass well before the critical limit is reached. Thus, choosing here a subcritical velocity for triggering the mechanical mass loss is not unrealistic. The star remains at the lower limit of the surface rotation for triggering the mechanical mass loss until the end of the MS phase (see Fig. \ref{fig:Vsurf_180Msol_z14_z06_ze5_z0}. After the MS phase, the star rapidly expands, the surface rotation becomes much lower than the limit and the mechanical mass loss stops. Other processes like line-driven winds or mass loss triggered by the continuum when some outer layers have supra Eddington luminosities become dominant.

The zero metallicity model has a similar evolution on the MS as the Z=10$^{-5}$ since it is not dominated by the line-driven mass loss. However, the Pop III model has not been pushed further in its evolution due to numerical problems linked to mechanical mass loss.

\subsection{Wolf-Rayet stars from VMS}

\begin{figure*}
    \centering
    \includegraphics[width=0.99\textwidth]{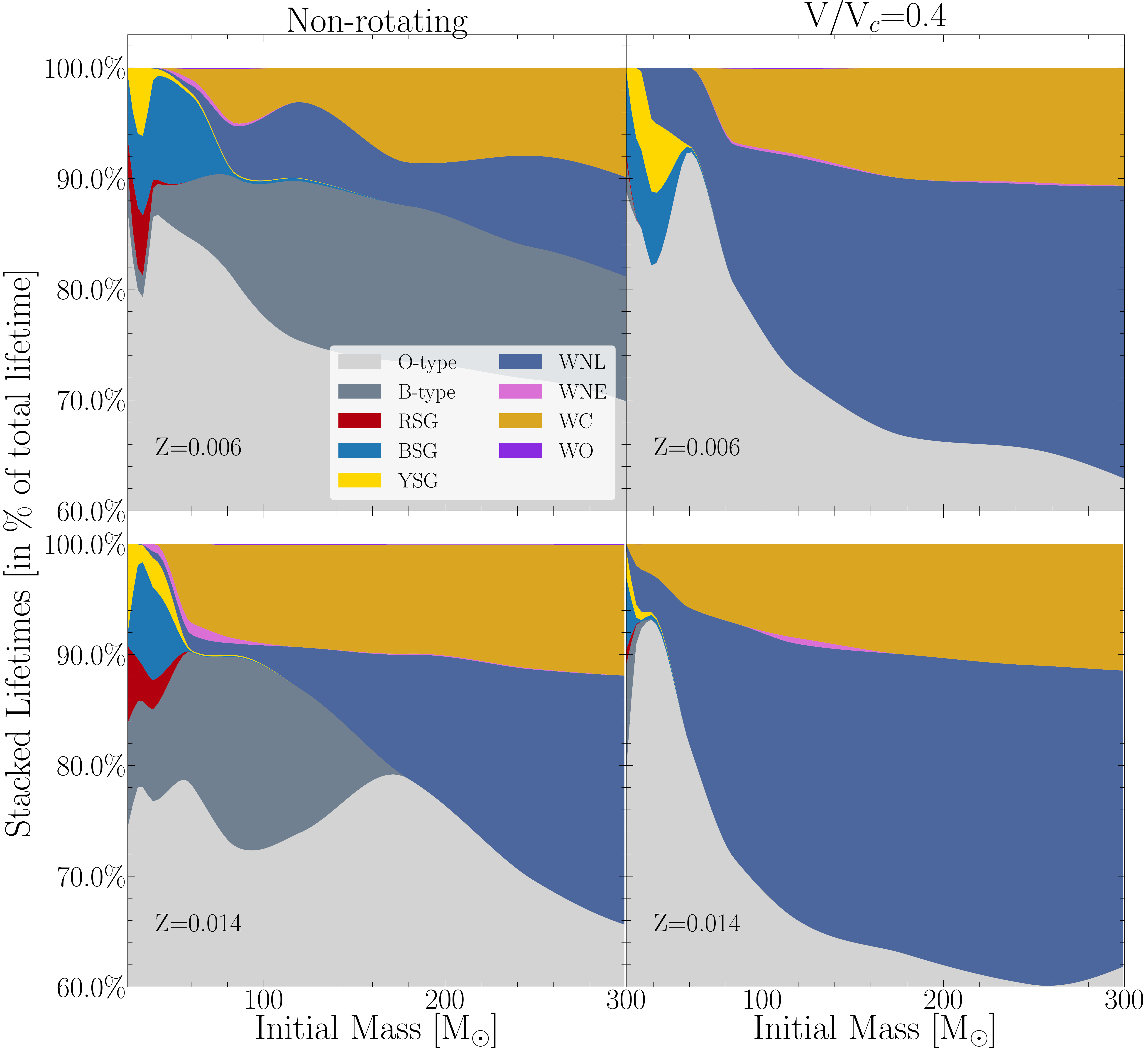}
    \caption{Stacked lifetimes as a percentage of the total lifetime for O-type stars, red supergiants (RSG), yellow supergiants (YSG) blue supergiants (BSG) and Wolf-Rayet types for non-rotating (left) and rotating models (right) at Z=0.006 (top) and Z=0.014 (bottom). }
    \label{fig:Stacked_lifetimes_percentage_z06}
\end{figure*}

\begin{figure*}[!h]
    \centering
    \includegraphics[width=0.93\textwidth]{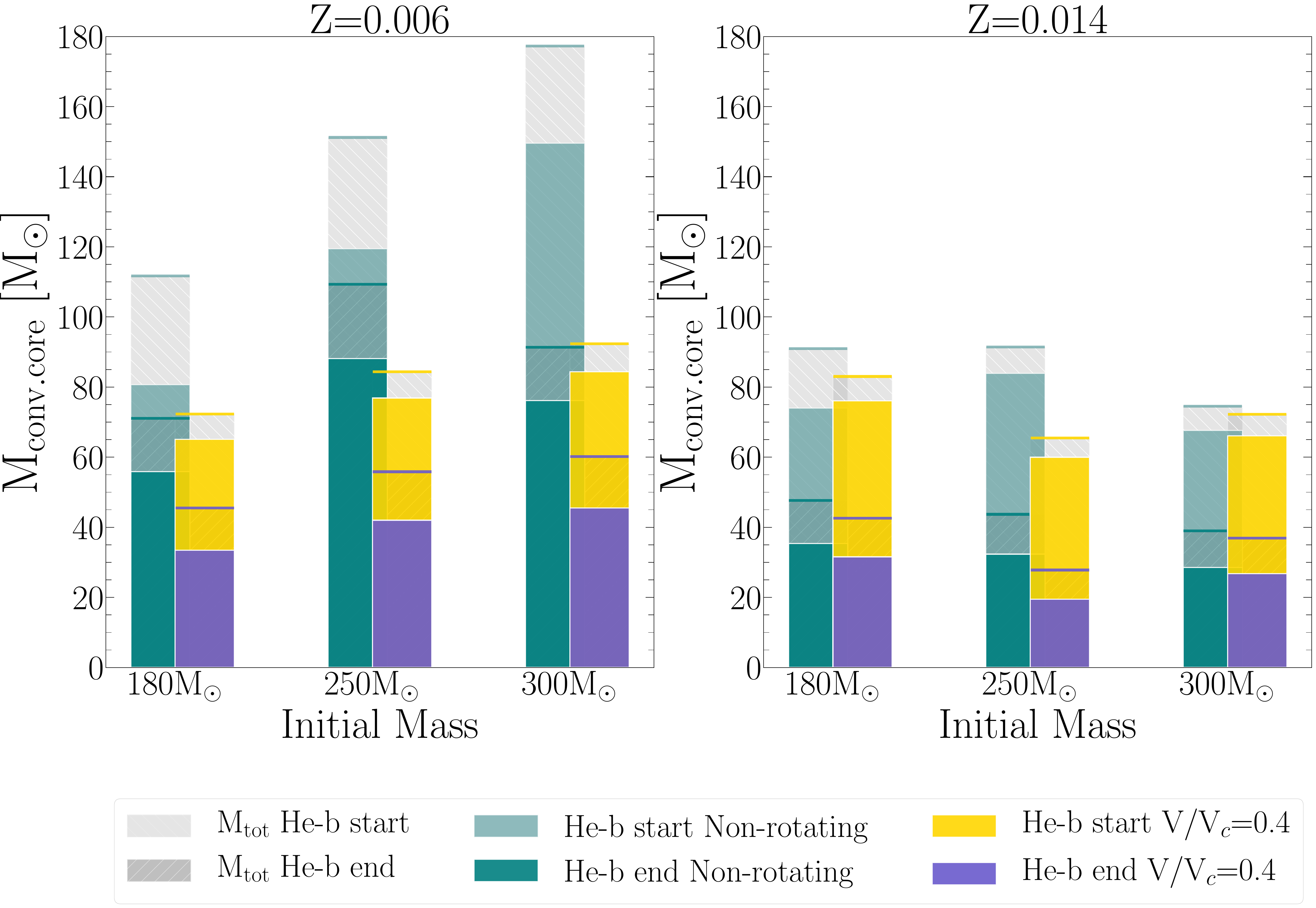}
    \caption{Evolution of the convective core mass of VMS stars at the beginning and at the end of the He-burning phase, for non-rotating and V/V$_c$=0.4 models at both Z=0.006 and Z=0.014. The light grey region corresponds to the total actual mass of the star at the considered evolutionary stage, and is emphasized by a line of the corresponding color.}
    \label{fig:conv_core_mass_histogra}
\end{figure*}

Wolf-Rayet (WR) stars are classified into different categories from their spectral features. 
As we do not predict the output spectra of our models, we cannot use here the same spectroscopic criteria as those used by spectroscopists to classify our models among the different WR sub-types.
Here  we have adopted different theoretical criteria for WR classification. The Wolf-Rayet phase in GENEC is assumed to begin when the model has an effective temperature larger than 10 000 K and a surface mass fraction of hydrogen X$^{^1\rm H}_s$ below 0.3. Assuming the star fulfills these WR conditions: 
\begin{itemize}
    \item the WNL phase is defined when X$^{^1\rm H}_s>~$10$^{-5}$
    \item the WNE phase when both X$^{^1\rm H}_s\leq~$10$^{-5}$ and the surface mass fraction of carbon X$^{^{12}\rm C}_s$ is smaller or equal to the one of nitrogen X$^{^{14}\rm N}_s$
    \item the WCO phase when X$^{^1\rm H}_s\leq~$10$^{-5}$ and X$^{^{12}\rm C}_s>~$X$^{^{14}\rm N}_s$
    \item the WO phase when it is a WCO with log(T$_{\rm eff})>~$5.25
    \item the WC phase when it is a WCO with log(T$_{\rm eff})\leq~$5.25
\end{itemize}
It is important to note that these classification criteria may actually result in differences with the ones obtained from spectroscopy \citep[see][]{Groh2014}. This is why the WO/WC criteria have been defined following the work of \citet{Groh2014}.

Figure \ref{fig:Stacked_lifetimes_percentage_z06} shows the stacked lifetimes of different spectral phases of massive stars as a percentage of the total lifetime of the concerned star. The lifetimes are displayed for non-rotating (left panels) and rotating at V/V$_c$=0.4 (right panels) models at Z=0.006 on the upper panels and Z=0.014 in the lower panels. 
The Z=0.006 models below 150 \Msol\ are coming from \citet{Eggenberger2021} and from \citet{Ekstrom2012} for the Z=0.014 ones. 

{Most of the life of massive stars is spent as O-type MS stars (note that the y-axis begins at 60\%).
The post-MS O-type star phase comprises between 10 and 40\% depending on the initial mass, metallicity and rotation. In general, this phase covers a larger fraction of the total lifetime when the initial mass, the metallicity, and/or the initial rotation increases. For example, the whole post O-type MS phase lasts about 27\% of the total lifetime of a 250\Msol\ non-rotating Z=0.006 model. Increasing the metallicity to Z=0.014 brings that fraction to about 30\%. The rotating models at Z=0.006 and 0.014 still increase that fraction to respectively 35 and 40\%. Except in the case of the non-rotating Z=0.006 model, where the fractions of the total lifetime  spent as a B-type, WNL and WC stars are more or less the same, for the other models, the post O-type MS star phase is only divided in two type of stars, either WNL (the longest ones covering about 2/3 of the remaining time) and the WC phase (one third).
This means that statistically, for single stars in a region with a constant star formation rate (for example during the last ten million years or so), there are in these last cases twice more chances to observe a WNL than a WC.}

\begin{figure}[!h]
    \centering
    \includegraphics[width=0.46\textwidth]{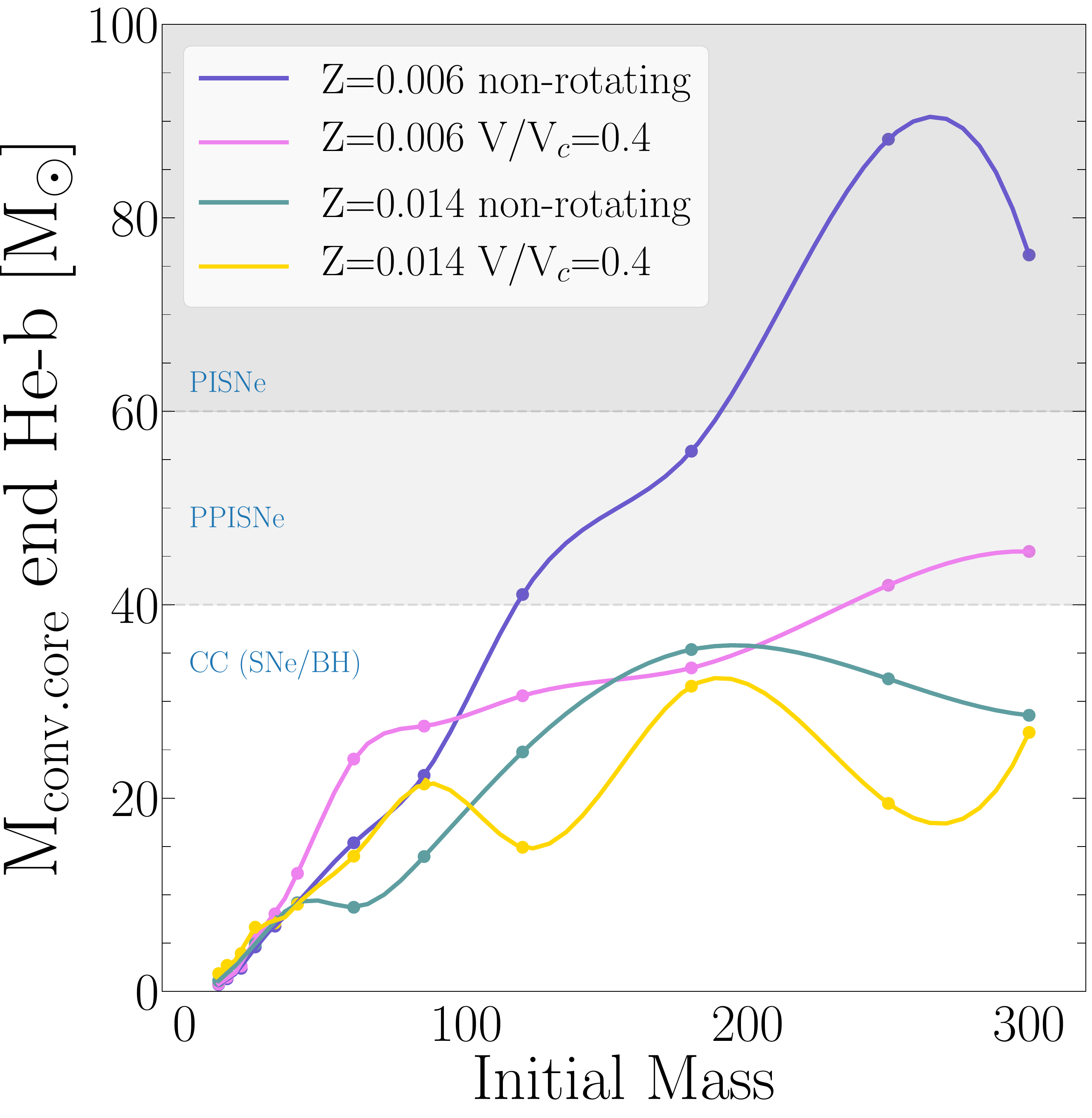}
    \caption{Convective core mass at the end of He-burning as a function of the initial mass for non-rotating and rotating stars at Z=0.006 and Z=0.014. The models with initial masses lower than 180\Msol\ are from \citet{Ekstrom2012} for Z=0.014 and \citet{Eggenberger2021} for Z=0.006, and share the same physic, albeit the differences described in Sect. \ref{Sect:Ingredients_of_models}. The range of CO core masses from \citet{Farmer2019} where (P)PISNe can occur is shown as an example to underline the very different potential final fate of VMS when considering rotation and metallicity effects.}
    \label{fig:limpairs}
\end{figure}

As is well known from previous models of rotating massive stars \citep[see e.g. Fig. 2 in][]{Georgy2012}, rotational mixing increases significantly the mass of the star that has a chemical composition enriched by H-burning products, hence the longer duration of the WN phase in rotating models than in non-rotating ones.

\subsection{Evolution of the core mass and final fate}

The CO core mass is an important quantity that affects the advanced stages of massive stars and especially their final fate. This quantity is highly dependent on the initial mass, but also on rotation and mass loss events.

Figure \ref{fig:conv_core_mass_histogra} shows the evolution of the convective core mass of VMS between the beginning and the end of the He-burning phase, for different rotation rates and metallicities. The convective core at the end of He-burning is a good proxy for the CO core mass. The figure also shows the actual total mass of the star at the different evolutionary stages shown on that figure. 

We see that for the masses considered in Fig. \ref{fig:conv_core_mass_histogra}, the total mass significantly decreases during the core He-burning phase and imposes the convective core to also decrease. 
We note that at the metallicity Z=0.006, the total mass of the non-rotating models at the beginning of the core He-burning phase increases with the initial mass, and the same for the convective core mass. This trend disappears at the end of the core He-burning phase. Typically, the total actual mass of the 300
M$_\odot$ model is smaller than that of the 250 M$_\odot$ one.
This is a consequence of the fact that the 300 M$_\odot$ entering into the core He-burning phase with a higher mass, loses more rapidly mass than the 250 M$_\odot$, allowing the model to lose a larger fraction of its mass by the end of that phase. This reflects the fact that the mass loss rates increase with the luminosity that itself increases with the actual mass of the star. The rotating models show much fewer differences between the different initial masses. They begin their core He-burning phase with similar actual mass and lose a similar amount of their mass during that phase. Rotational mixing, as shown above, makes the star enter at an early evolutionary stage in the WR phase. From that stage, the mass loss scales with the luminosity and hence the mass, making the models converge towards similar masses. Indeed, as explained above, by starting with more mass, the high luminosity will produce stronger mass losses and thus causes the model to reach a similar mass as models starting with a lower mass.

At solar metallicity, the differences between the non-rotating and rotating models are much smaller than at Z=0.006. The strong winds dominate here, leaving little room for the effects of rotational mixing.



Very often in the literature, the mass of the He core or that of the CO core at the end of He-burning is used to determine whether the star will explode as PISN or PPISNe \citep[see for example][]{Farmer2019,Costa2021}. As we have seen, the convective core mass at the end of He-burning is a good proxy for the CO core mass.
In Fig.~\ref{fig:limpairs}, the CO core masses limits given by \citet{Farmer2019} are indicated. We see that according to this criterion, the only two models that will explode as a pair-instability supernova would be the 250 and 300 M$_\odot$ non-rotating models at Z=0.006. 
The inferior mass limit would be 200 M$_\odot$ for entering into the domain of PISNe.
The result is different from the one by \citet{Yusof2013} (see therein their Fig.~18). 
They find that for the non-rotating Z=0.006, the inferior mass limit for the non-rotating models was found to be around 300 M$_\odot$. This is likely due to the absence of models between 150 and 500\Msol\ in the \citet{Yusof2013} study. Indeed, we see that the final mass has a peak for an initial mass around 260\Msol\ in this study, so it is likely that the \citet{Yusof2013} study missed that peak.
For rotating models at Z=0.006, the 250 and 300\Msol\ models are expected to undergo PPISNe.

{We note also that the models for Z=0.014 are expected to avoid both Pulsational Pair instability phase and Pair instability explosions for all the models computed here, thus at least for models up to an initial mass of 300 M$_\odot$.  In case those models end their lives as black holes engulfing the whole mass that the star has retained until the final core collapse, black hole masses up to 30 - 45 M$_\odot$ could be formed. Still higher mass black holes (up to about 60 M$_\odot$) could be formed from the non-rotating Z=0.006 models if the mass loss induced by the pulsational pair instability would be smaller than a few solar masses.}

In a paper in preparation, we will present a more in-depth study of the final fate of VMS where we will study both core masses criteria and an average value for the first adiabatic exponent over the whole mass \citep[see][]{Stothers1999,Marchant2019,Costa2021}. We will discuss whether indeed those models that should explode as a (P)PISNe according to the core mass criterion show a sufficiently large fraction of their total mass unstable to trigger the explosion mechanism (Martinet et al. in preparation).

\section{Comparison with previous models}

\begin{figure*}
    \centering
    \includegraphics[width=0.47\textwidth]{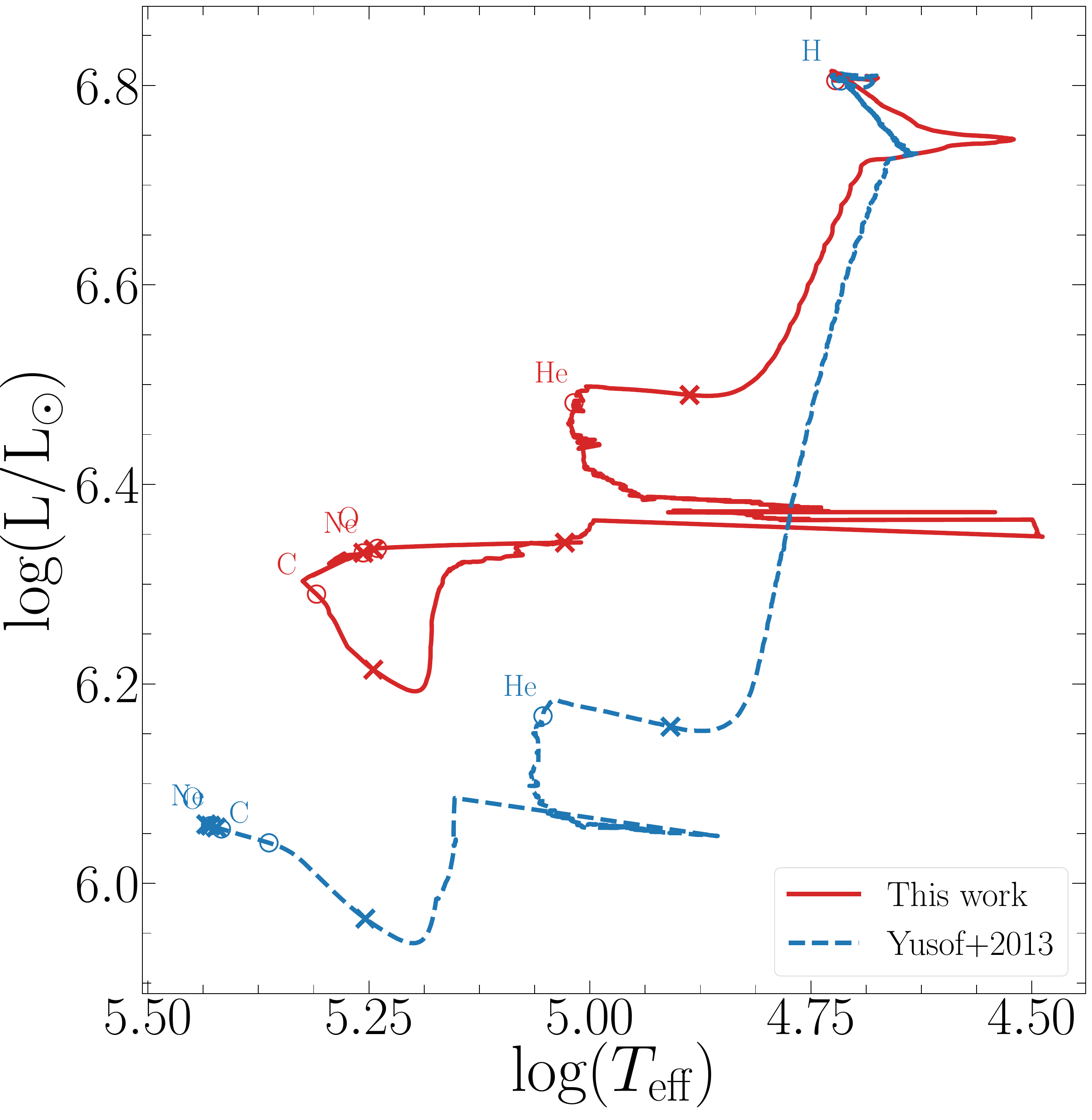}
    \includegraphics[width=0.47\textwidth]{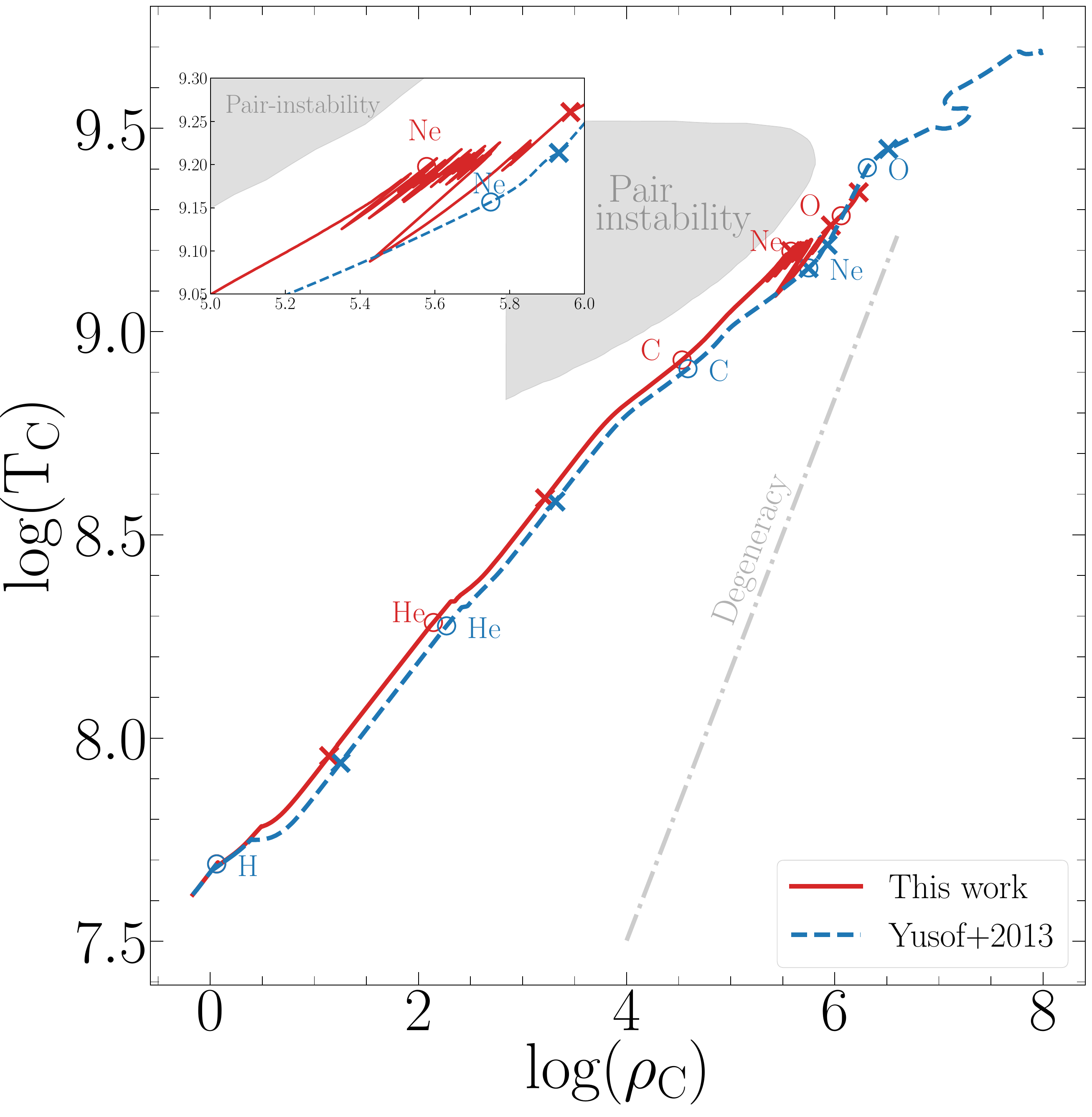}
    \caption{\textit{Left:} HRD of a rotating 300\Msol\ at Z=0.014 from the work of \citet{Yusof2013} and from this work. The beginning of each burning phase (defined when the central abundance of the concerned element decreased by 1\%) and the end of it (defined by the central abundance being lower than 10$^{-5}$) are displayed respectively as open circles and crosses. \textit{Right:} Central temperature as a function of the central density for the same models. A zoom can be seen in the upper left displaying the region where the model starts undergoing pair-creation inside the core, leading the model with the new EOS taking this effect into account to contract and expand rapidly as the pair-creation reduces the radiative pressure.}
    \label{fig:HRD_rhoT_Martinet_vs_Yusof}
\end{figure*}



Figure \ref{fig:HRD_rhoT_Martinet_vs_Yusof} presents the evolution of a rotating (V/V$_c$=0.4) 300 \Msol\ at Z=0.014 computed with GENEC, from the computations of \citet{Yusof2013}, and the other one from this work. The main differences between these two computations are mainly coming from the improvement to the code, namely the treatment of conservation of angular momentum in the envelope \citep[see][]{Ekstrom2012}, and the new EOS we have been introducing to take into account the pair-creation effects. As we also have seen in Sect. \ref{Sect:Ingredients_of_models}, these new VMS models are computed with the Ledoux criterion and an increased overshoot $\alpha_{\rm ov}$=0.2, while \citet{Yusof2013} models are computed with the Schwarzschild criterion and $\alpha_{\rm ov}$=0.1. The left panel of Fig. \ref{fig:HRD_rhoT_Martinet_vs_Yusof} shows the HRD of these two models, with the beginning and end of the different burning phases depicted by circles and crosses. 

The beginning of the MS is very similar for both models, however slight discrepancies appear at the surface during the rest of the MS evolution. This is due to the different treatment of the angular momentum conservation at the surface, combined with the very large mass loss rates occurring for such massive stars at solar metallicity. The larger overshooting parameter does not induce a meaningful increase of the core size during the MS due to VMS having already very large convective cores. The total mass of both models during most of the MS is similar because mass loss dominates its evolution. However, at the end of the H-burning, a large mass loss occurs for both models, leading however to larger quantities lost in the \citet{Yusof2013} models. This leads our models to be 40\% larger in mass at the end of the MS. 

The He-burning phase occurs at higher luminosity and goes through a slight evolution to the red for the model of this work. The leading consequence is that the VMS will undergo slightly larger mass loss rates during the middle of the He-burning when going to the redder part of the HRD. {This reduces the difference in the actual mass between our model and the one by \citet{Yusof2013} from 40\% at the end of the MS phase to
35\% at the end of the core He-burning phase.}
Interestingly, the evolution from the C-burning phase is slightly different, with \citet{Yusof2013} model continuing to higher effective temperature while this work model evolves back to the red. To explain this feature, we need to now go to the right panel of Fig. \ref{fig:HRD_rhoT_Martinet_vs_Yusof} where the central temperature of the star is depicted as a function of the central density. Once again we see that in early stages, the central conditions are very similar, with a slightly higher temperature in the core in the new model ({likely due to the larger overshoot parameter}). The very interesting features appear at the end of the C-burning phase and are displayed as a zoom in the upper-right part of the diagram. The oscillations of the central conditions are in fact a direct effect of the pair production inside the star. Indeed, while the center is out of the pair-instability zone, a part of the core slightly off-centered is crossing this instability region. The production of pair inside these parts of the star results in destabilizing this region of the core\footnote{Note that the models are always by construction here at hydrostatic equilibrium. The oscillation seen here is likely due to the fact that pair-creation remove radiative support and thus triggers contraction. At its turn, contraction leads to larger nuclear energy production  that triggers expansion.}.

\section{Comparison with observed VMS in the LMC}

\begin{figure*}
    \centering
    \includegraphics[width=0.99\textwidth]{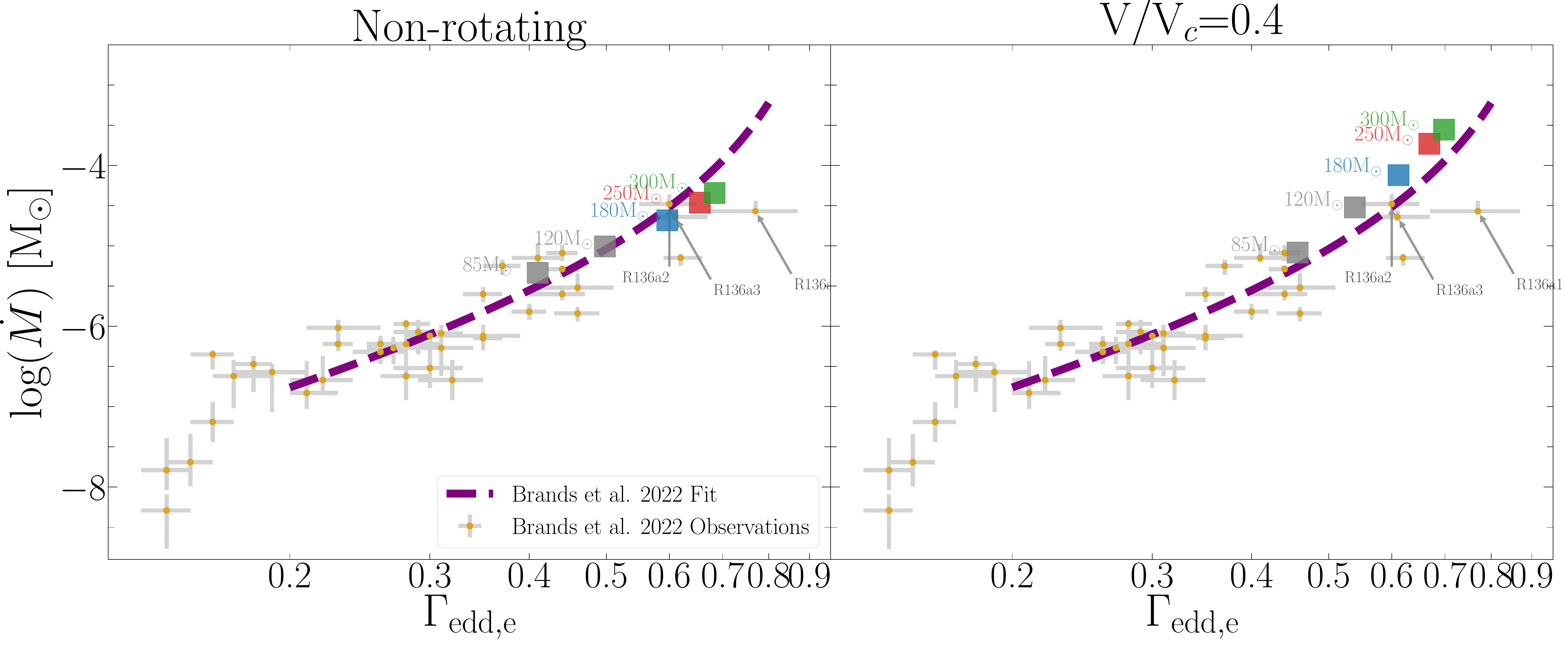}
    \caption{Time-averaged mass loss rates over the Main-Sequence as a function of the Eddington parameter $\Gamma_{\rm edd,e}$. The models of 85-120\Msol\ are from \citet{Eggenberger2021}, and the 180-300\Msol\ from this work. The observations of \citet{Brands2022} of the R136 components are displayed in gold. The fit they obtained is displayed in purple. The three most massive components of the cluster (R136a1/2/3) are pinpointed by arrows.}
    \label{fig:Mdot_Gamma_Edd_Brands2022}
\end{figure*}
\begin{figure*}
    \centering
    \includegraphics[width=0.99\textwidth]{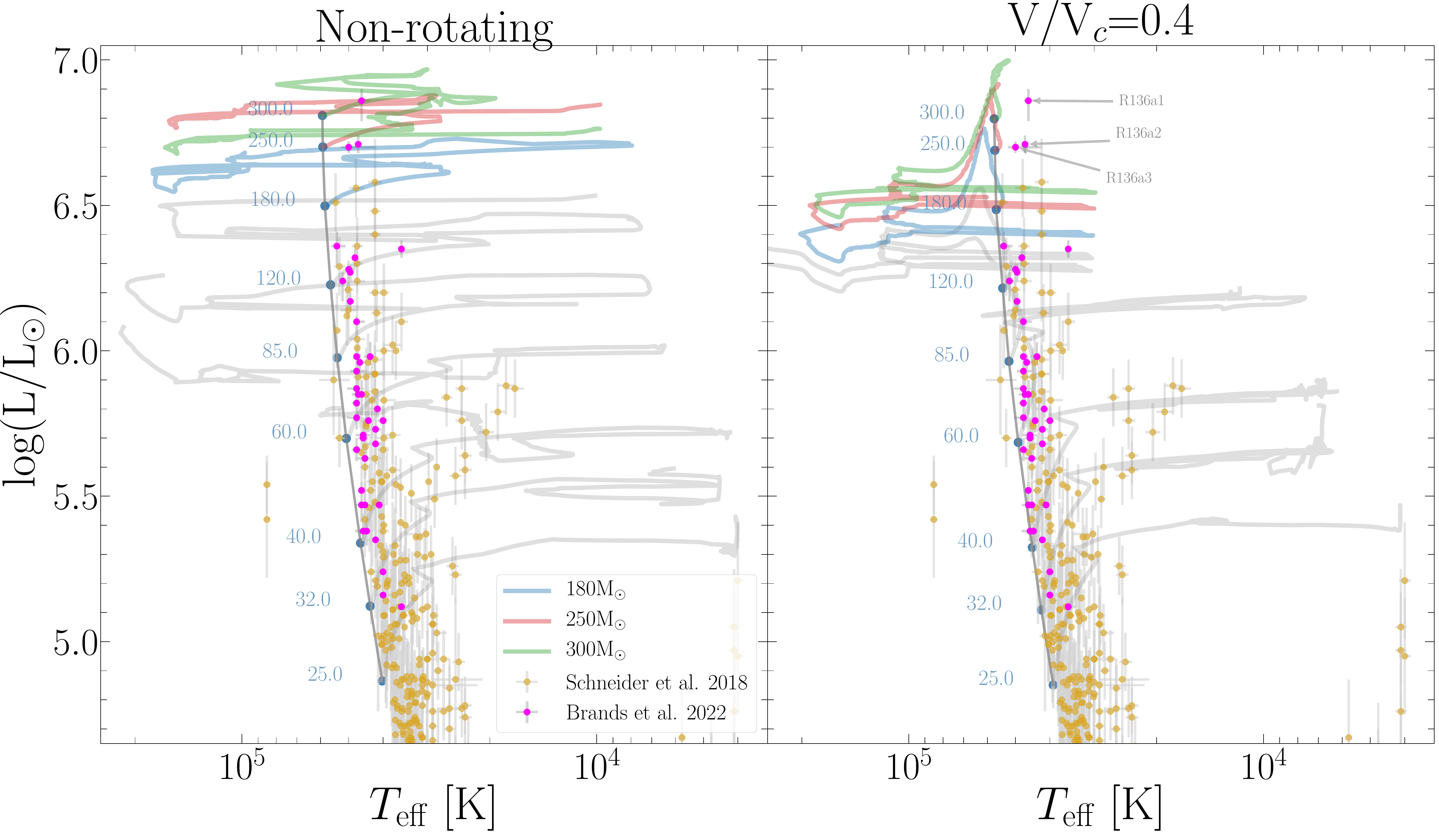}
    \caption{HRD of the Tarantula Nebula from the compilation of \citet{Schneider2018} \citep[regrouping results from][]{RamirezAgudelo2017,SabinSanjulian2014,SabinSanjulian2017,McEvoy2015,Bestenlehner2014} and the R136 cluster members from the work of \citet{Brands2022}. The GENEC tracks are overplotted in grey, with color emphasis on the VMS models computed here, and the ZAMS is drawn in dark grey. The HRD is plotted for non-rotating (\textit{left}) and rotating at V/V$_c$=0.4 (\textit{right}) models. Their effective temperatures are not corrected for the optical thickness of the wind. The three most massive components of the cluster (R136a1/2/3) are pinpointed by arrows on the right panel.}
    \label{fig:HRD_schneider2018}
\end{figure*}
The Tarantula Nebula in the Large Magellanic Cloud hosts one particular young cluster, known as R136, hosting itself very high mass components. 
Recent work \citep{Bestenlehner2020,Brands2022} confirmed these results with new VLT-FLAMES and HST/STIS optic/spectroscopic measurements and provide interesting stellar characteristics to the most massive stars observed. 
With the new VMS models computed here, we can confront our results to these new observations. We explore here the impact of the physics we have included inside our models on the observable, and compare the results to the ones provided by the most recent observations.

As we have seen in the previous section, the evolution of VMS at high metallicity is dominated by mass loss. The prescriptions used in our models are coming from both theoretical and empirical results obtained from lower mass O-type stars, and are highly uncertain, especially when getting close to the Eddington limit \citep[see the review of][]{Vink2021}. We confront in Fig. \ref{fig:Mdot_Gamma_Edd_Brands2022} our models to the observed mass loss rates obtained by \citet{Brands2022} and their corresponding fit using the prescription proposed by \citet{Bestenlehner2020}. The mass loss rates are plotted as a function of the Eddington parameter (with $\Gamma_{\rm edd,e}$=1 being the Eddington limit). The time-averaged mass loss of our models over the Main-Sequence is shown by colored rectangles. The time-averaged mass loss rates of our non-rotating VMS models with initial masses equal or above 180 M$_\odot$ are below the fitting line by at most 0.3 dex (a facor two). The contrary for our rotating models that show time-averaged mass loss rates at most about 0.3 dex above the fitting line. At least, this shows that although we did not account for the most recent developments of the mass loss rates in the present models, the fact that
the non-rotating and rotating models more or less frame the fitting curve by \citet{Brands2022} with
a factor of at most a factor 2 is rather encouraging. Indeed, the scatter of the colored squares is smaller than the scatter of the individual mass loss rate determinations.


Figure \ref{fig:HRD_schneider2018} presents the HRD of the Tarantula Nebula, a very active star-forming region in the LMC. The compilation of \citet{Schneider2018} is displayed in yellow, while the components of the R136 cluster, obtained by \citet{Brands2022}, are displayed in magenta. The tracks used here are the Z=0.006 GENEC tracks from \citet{Eggenberger2021} and the VMS tracks we computed for this work with the new EOS. We can see that this cluster hosts several high-mass components. If we look at the evolution of our VMS models, three stars inside this cluster are above M$_{\rm ini}$=200\Msol\ (R136a1/2/3). 
At first glance, one would favor the non-rotating models for these VMS, covering a large range of effective temperatures on the MS, while the rotating models, as we have seen in Sect. \ref{sect:impact_of_metallicity}, evolve quasi-chemically homogeneously and stray directly to the bluer part of the HRD (also true for models rotating at V/V$_c$=0.2 not shown here), hence apparently incompatible with the observed effective temperatures. The problem is however more complicated, the VMS undergo very large mass loss already during the MS. This means, in fact, that the star is embedded inside a "nebula" of very thick wind. When observing such stars, we observe in fact this thick wind, and this blurs the determination of the effective temperature of the surface of the star. Indeed, with the wind being much cooler than the surface of the star, we would obtain a lower effective temperature to be observed if we took this effect into account. We computed corrected effective temperatures and find that both non-rotating and rotating models are compatible with the observations (the corrected effective temperatures go down to 35000K, 6000K below the lowest limit of the observed T$_{\rm eff}$ of VMS see the details of how this correction has been computed in \citet{Schaller1992}).

\begin{figure}
    \centering
    \includegraphics[width=0.48\textwidth]{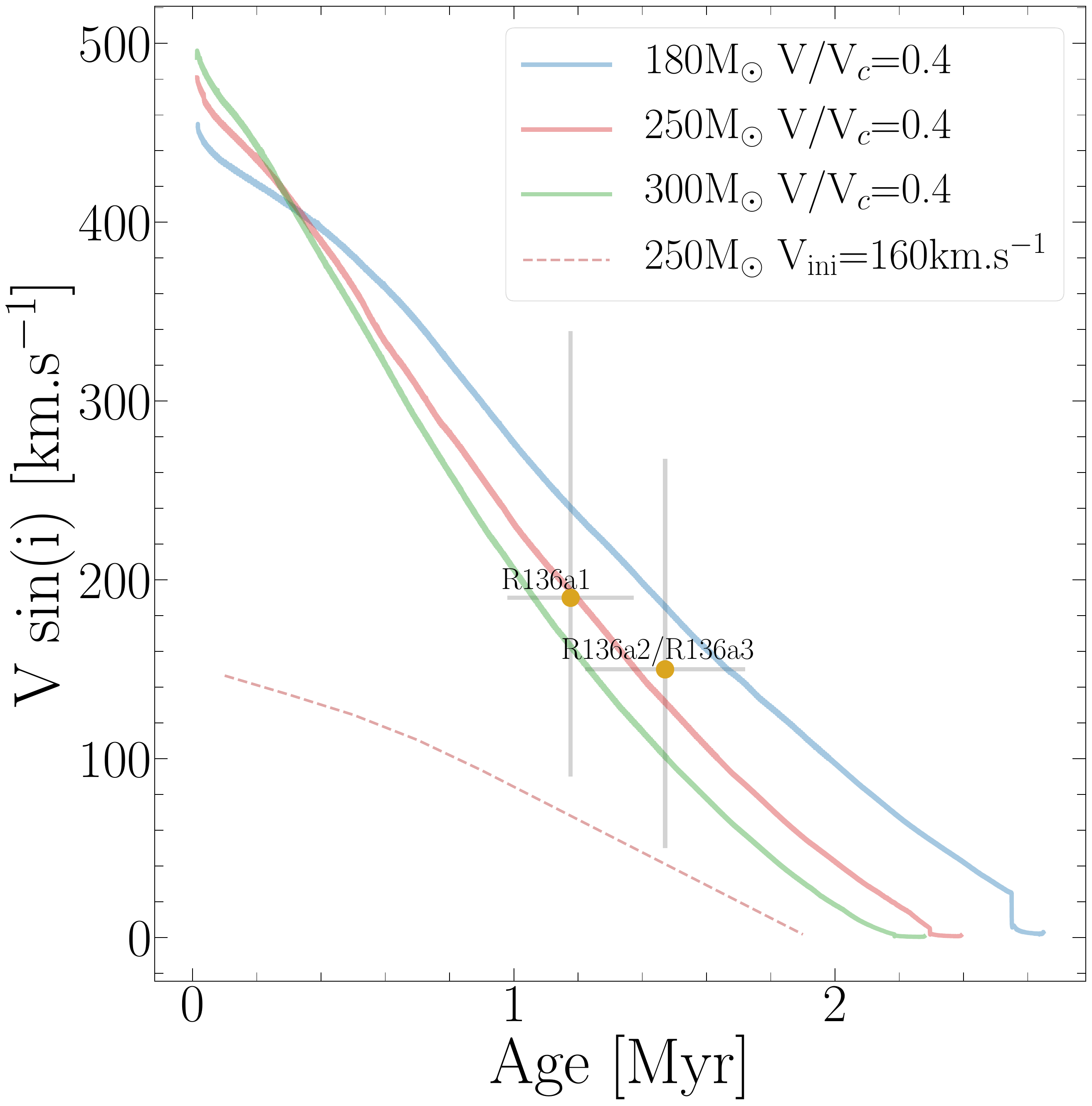}
    \caption{Evolution of the surface velocity of VMS models. An additional model with slower initial rotation rate has been computed for this plot. The observed velocities of the three most massive components of R136 have been overplotted (R136a2 and R136a3 have similar velocities obtained). The uncertainties on the velocity are taken from the maximum macro-turbulent velocity observed for stars close to the Eddington limit \citep{SimonDiaz2017} for the lower limit, and from the maximum velocities obtained from angle of view effect for the upper limit.}
    \label{fig:Vsini_VMS_observed}
\end{figure}

\begin{figure}
    \centering
    \includegraphics[width=0.48\textwidth]{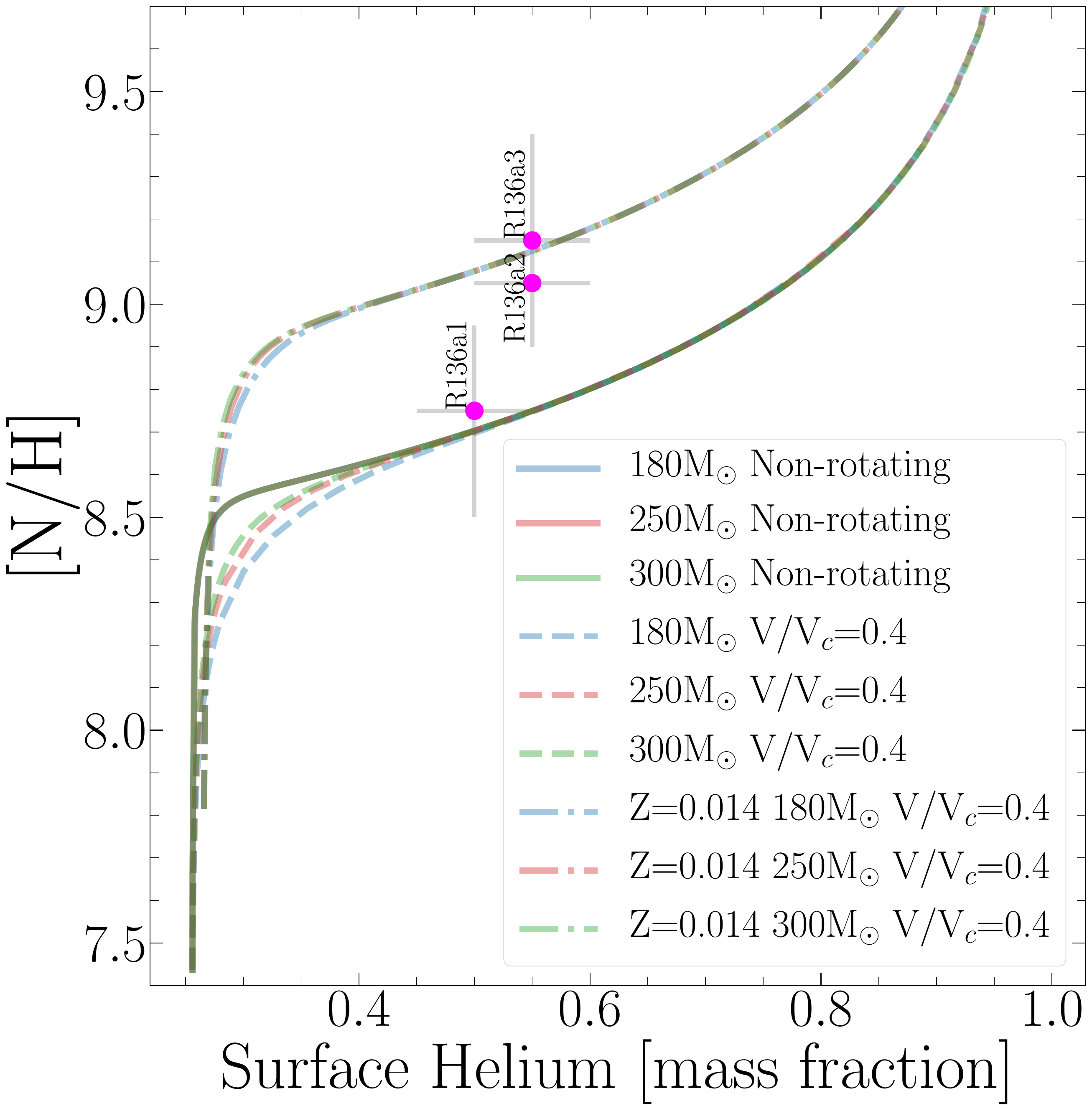}
    \caption{Surface enrichment of Nitrogen over Hydrogen as a function of the surface Helium. The observed values of [N/H] are from \citet{Brands2022}, and the surface Helium fraction from \citet{Bestenlehner2020}.}
    \label{fig:NHrel_VMS_observed}
\end{figure}

Figure \ref{fig:Vsini_VMS_observed} shows the values of $V\sin i$ obtained by \citet{Bestenlehner2020} overplotted over the tracks of VMS models. From the spectroscopic studies, the surface velocities observed do not differentiate the surface velocity from the macro-turbulent velocity. This means that the lower limit of the surface velocity would be when the macro-turbulent is at a maximum. We take the maximum macro-turbulent velocities obtained by \citet{SimonDiaz2017} for stars close to the Eddington limit to obtain the lower displayed limit. The upper limit is obtained from the angle of view effects. 

The theoretical models (see the continuous lines in Fig. \ref{fig:Vsini_VMS_observed}) shows that the surface velocity decreases rapidly as a function of time as a result of the large amount of angular momentum taken away by stellar winds.

To obtain that the theoretical model goes through the region covered by the error bars given by \citet{Bestenlehner2020} at this stage of the MS, we need quite high rotation rates already at the ZAMS. Actually, since, as recalled above, the $V\sin i$ value is a lower limit to the true surface velocity, even larger initial rotations than those shown for the models in Fig. \ref{fig:Vsini_VMS_observed} could be considered.

A 250\Msol\ model with lower rotation rate has been computed to show that the mass loss still dominates the surface velocity evolution and would not be able to reproduce the high velocities observed for these VMS. We can wonder whether assuming a solid body rotation as would do a model accounting for the impact of an internal magnetic field, would allow to start from  a lower initial rotation rate. This will be explored in the future. However, we can note that despite the fact that we considered non-magnetic models here, since these stars host very massive convective core they have a solid body rotation anyway in a very large part of the star. We suspect therefore that including magnetic fields in our models would have here a modest effect.
(see discussion in Sect. \ref{sect:conclusion_physical_ingredients}).    

Figure \ref{fig:NHrel_VMS_observed} shows the observed surface enrichment of nitrogen as a function of the observed surface helium, respectively from \citet{Brands2022} and \citet{Bestenlehner2020}. The non-rotating (in solid lines) and the rotating (in dashed lines) VMS models are overplotted. Interestingly, the evolution of nitrogen at the surface is very similar between the non-rotating and rotating models. This is because the models quickly reach (before the surface Helium fraction reaches 0.4) the maximum quantity of nitrogen produced from the CNO cycle in these stars. This maximum value depends solely on the initial abundances of CNO. Hence, while the most massive component R136a1 is well reproduced by the models, there is no possibility for these models to explain the very high enrichment value of nitrogen of the two other VMS (R136a2/a3). Understanding these discrepancies is difficult. A first possibility could be that the initial abundances and in particular the initial content of CNO elements could be different in the different stars.
We can see that the Z=0.014 models are able to reproduce such values due to the larger initial abundances of CNO. 
Can we imagine that this region can host populations with different metallicities?
This would imply, in a self-enrichment scenario, multiple stellar generations born from material differently enriched in metals by previous generations of stars.
There are massive young clusters that show evidence for multiple generations of stars, such as the Orion nebula cluster with potentially generations with an age separation of less than 1 Myr \citep{Beccari2017}. The Tarantula nebula is a much more massive star-forming region than Orion, even more susceptible to having hosted multiple generations. Interestingly,
a scenario where the (re)-collapse of the cluster allows the birth of a second generation of stars has been invoked by 
\citet{Dominguez2022} to explain the observational characteristics of the central region of 30 Doradus. 
A second possibility to explain the result shown in Fig.~\ref{fig:NHrel_VMS_observed} is that the error bars have been underestimated. The error bars should be significantly increased to allow tracks at a given fixed metallicity to cross 
them. This appears not very realistic.
Third, we can wonder if the past history of the star, such as binary interaction or early merger, could be invoked to explain the discrepancy in nitrogen surface abundances at so similar values for the surface helium abundance. Of course, some modeling of the evolution of multiple systems is required to answer such a question.

\section{Discussion and conclusions}

\subsection{Impact of changing physical ingredients of the models}
\label{sect:conclusion_physical_ingredients}

Changes of some physical ingredients as the criterion for convection, the overshooting, the expression for the diffusion coefficient in rotating models, or the change of the EOS have non-negligible impacts on the outputs of the stellar models as 
discussed in Sect.~4, where we performed a comparison with the models by \citet{Yusof2013} that used, otherwise also GENEC as the stellar evolution code.

For solar metallicity, models are dominated by mass loss, hence most of the differences come from 
the changes in angular momentum conservation and the slightly different evolution in the HRD. While non-rotating models are quite similar, rotating ones lead to final masses up to 42\% larger than previous models. At Z=0.006, the differences are smaller due to less mass loss occurring. For the rotating models, the final masses are up to 12\% larger than in \citet{Yusof2013} models.

We can wonder what would be the impact of changing the angular momentum transport by considering a more efficient process, such as the one proposed in the Taylor-Spruit (TS) dynamo \citep{Spruit2002,Mader&Meynet2004TaylorSpruit,Eggenberger2010}. The present VMS rotating models at V/V$_c$=0.4 are almost rotating as a solid body at the beginning of the MS due to their very large convective core. It is also the case for slower rotators such as the Z=0.006 180 \Msol\ rotating at V/V$_c\simeq$0.2 presented in Fig. \ref{fig:Vsini_VMS_observed}. The rotation of the envelope and the core stays coupled in fact up until the end of the MS for models at Z=0.006, when the mass loss becomes prevalent. A more efficient transport such as the TS dynamo would not change much the rotation profile in that case. For models at Z=0.014, differential rotation between envelope and core sets in from the middle of the MS, due to higher mass loss rates. In this case, a more efficient transport of angular momentum might produce stars with faster surface rotation.

{VMS are highly sensitive to mass loss, especially at high metallicities. We chose for this work to keep the same physical ingredients that were used for all the grids published so far \citep{Ekstrom2012,Georgy2013,Groh2019,Murphy2021,Eggenberger2021,Yusof2022}. In doing so, it allows us to discuss  how varying the initial mass, metallicity and rotation impact the outputs. While the mass loss prescription scheme we use here is used in many recent grids for massive stars \citep[see for example][]{Fragos2023,SimazBunzel2023,Jiang2023,Brinkman2023,Song2023}, recent works presented new predictions for mass loss in massive stars \citep[such as][]{Bjorklund2022,GormazMatamala2023,Yang2023}. \citet{Bjorklund2022} predict that O-stars mass loss rates are lower by about a factor 3 than the rates typically used in stellar-evolution calculations, however, differences decrease with increasing luminosity and temperature (\textit{i.e.} less difference for VMS and their high luminosity). Moreover, they find that the remaining key uncertainty regarding these predictions concerns unsteady mass loss for very high-luminosity stars close to the Eddington limit, a major mass loss component due to VMS being very close to the Eddington limit (as we have seen in Fig. \ref{fig:Mdot_Gamma_Edd_Brands2022}). 
For RSG mass loss rates, \citet{Massey2023} discussed the relevance of the prescriptions used in this work by comparing the predicted and observed luminosity function of red supergiants, finding a good agreement. Still, \citet{Yang2023} present a new prescription for RSG mass loss rates from a large spectroscopic survey in the Small Magellanic Cloud, finding that it may provide a more accurate relation at the cool and luminous region of the HR diagram at low metallicity compared to previous studies.
}
{The mass loss rates prescriptions we use at Z=0.006, although not considering the effects discussed in \citet{Bestenlehner2020}, provide a time-averaged mass loss rate during the Main-Sequence that is underestimated by a factor of up to 2 for the non-rotating models and overestimated by a factor up to 2 for the rotating models. The scatter is reasonably small when compared to the scatter of the individual mass loss rate determinations used to establish the fitting mass loss rate prescription.}
It is however uncertain how our current prescriptions reproduce mass loss rate at solar metallicity, and if we under- or over-estimate the quantity of mass lost by solar metallicity VMS \citep{Besten2014,Vink2021}. 
For the low metallicity case, while they undergo almost no mass loss during the MS, their mass loss close to the Eddington limit is very uncertain and might have an important impact on their evolution in the later stages \citep{Sander2020}.

{Studying the impact of recent mass loss rate prescriptions on VMS is crucial \citep[see][]{Sabhahit2022,Sabhahit2023,Higgins2023}, but updating important physic such as chemical mixing and transport of angular momentum would be as necessary. Thus, new generations of stellar models should not only update the mass loss rates but also the physics of many other physical ingredients. This is beyond the aim of the present work and will be the topic of future works. }
\subsection{Synthesis of the main results and future perspectives}

We computed a grid of VMS from Population III to solar metallicity with and without rotation. We showed that the low metallicity models are highly sensitive to rotation, while the evolution of higher metallicity models is dominated by mass loss effects.  The mass loss affects strongly their surface velocity evolution, quickly at high metallicity while reaching the critical velocity for low metallicity models. The comparison to observed VMS in the LMC showed the mass loss prescriptions used for these non-rotating/rotating models are slightly under-/over-estimating compared to the observed mass loss rates. According to the treatment of rotation in our models, observed VMS need a high initial velocity to be able to account for their observed surface velocity. The surface enrichment of these observed VMS is difficult to explain with only one initial composition and could suggest multiple populations in the R136 cluster or maybe a more complex scenario involving multiple star interactions.

The tremendous dependence of the evolution of VMS upon the physic input will have of course a crucial impact on the final fate and the nature of the remnants that such stars can produce. As shown in multiple studies \citep{Woosley2021,Costa2021,Marchant2020,Farmer2019}, the final fate of VMS is strongly linked to the CO core mass at the end of He-burning. This work provides a new illustration of the fact that rotation, mass loss, and other physic inputs play an important role in the CO core mass value, and could be essential to understand the final fate and the maximum mass of the remnants produced by VMS. {A first discussion shows that at a metallicity typical of R136, only the non- or slowly rotating VMS models may produce Pair Instability supernovae. The most massive black holes that could be formed then are expected to be less massive than about 60 M$_\odot$.}
In a dedicated follow-up paper, we will discuss the impact of the different physics introduced here on the final fate of the stars. Thanks to the addition of an equation of state following the pair production inside the star, we can now track the evolution of the pair-instability inside VMS. This will allow us to determine for the whole grid of models which ones are either undergoing core collapse, pulsational-, pair-instability supernovae or direct collapse through pair-instability.

\begin{acknowledgements}
SM has received support from the SNS project number 200020-205154, the European Union’s Horizon 2020 research and innovation program under grant agreement No 101008324 (ChETEC-INFRA) and by the Fonds de la Recherche Scientifique (FNRS, Belgium) and the Research Foundation Flanders (FWO, Belgium) under the EOS Project nr O022818F and O000422.
GM, SE, CG and have received funding from the European Research Council (ERC) under the European Union's Horizon 2020 research and innovation program (grant agreement No 833925, project STAREX).
RH acknowledges support from the World Premier International Research Centre Initiative (WPI Initiative, MEXT, Japan), STFC UK, the European Union’s Horizon 2020 research and innovation program under grant agreement No 101008324 (ChETEC-INFRA) and the IReNA AccelNet Network of Networks, supported by the National Science Foundation under Grant No. OISE-1927130. 
\end{acknowledgements}

\bibliographystyle{aa} 
\bibliography{Thesis_2} 

\begin{thebibliography}{104}
\expandafter\ifx\csname natexlab\endcsname\relax\def\natexlab#1{#1}\fi

\bibitem[{{Abel} {et~al.}(2002){Abel}, {Bryan}, \& {Norman}}]{Abel2002}
{Abel}, T., {Bryan}, G.~L., \& {Norman}, M.~L. 2002, Science, 295, 93

\bibitem[{{Bastian} \& {Lardo}(2018)}]{Bastian2018}
{Bastian}, N. \& {Lardo}, C. 2018, \araa, 56, 83

\bibitem[{{Beccari} {et~al.}(2017){Beccari}, {Petr-Gotzens}, {Boffin},
  {Romaniello}, {Fedele}, {Carraro}, {De Marchi}, {de Wit}, {Drew}, {Kalari},
  {Manara}, {Martin}, {Mieske}, {Panagia}, {Testi}, {Vink}, {Walsh}, \&
  {Wright}}]{Beccari2017}
{Beccari}, G., {Petr-Gotzens}, M.~G., {Boffin}, H.~M.~J., {et~al.} 2017, \aap,
  604, A22

\bibitem[{{Becker} \& {Bolton}(2013)}]{Becker2013}
{Becker}, G.~D. \& {Bolton}, J.~S. 2013, \mnras, 436, 1023

\bibitem[{{Bestenlehner} {et~al.}(2020){Bestenlehner}, {Crowther},
  {Caballero-Nieves}, {Schneider}, {Sim{\'o}n-D{\'\i}az}, {Brands}, {de Koter},
  {Gr{\"a}fener}, {Herrero}, {Langer}, {Lennon}, {Ma{\'\i}z Apell{\'a}niz},
  {Puls}, \& {Vink}}]{Bestenlehner2020}
{Bestenlehner}, J.~M., {Crowther}, P.~A., {Caballero-Nieves}, S.~M., {et~al.}
  2020, \mnras, 499, 1918

\bibitem[{{Bestenlehner} {et~al.}(2014{\natexlab{a}}){Bestenlehner},
  {Gr{\"a}fener}, {Vink}, {Najarro}, {de Koter}, {Sana}, {Evans}, {Crowther},
  {H{\'e}nault-Brunet}, {Herrero}, {Langer}, {Schneider},
  {Sim{\'o}n-D{\'\i}az}, {Taylor}, \& {Walborn}}]{Bestenlehner2014}
{Bestenlehner}, J.~M., {Gr{\"a}fener}, G., {Vink}, J.~S., {et~al.}
  2014{\natexlab{a}}, \aap, 570, A38

\bibitem[{{Bestenlehner} {et~al.}(2014{\natexlab{b}}){Bestenlehner},
  {Gr{\"a}fener}, {Vink}, {Najarro}, {de Koter}, {Sana}, {Evans}, {Crowther},
  {H{\'e}nault-Brunet}, {Herrero}, {Langer}, {Schneider},
  {Sim{\'o}n-D{\'\i}az}, {Taylor}, \& {Walborn}}]{Besten2014}
{Bestenlehner}, J.~M., {Gr{\"a}fener}, G., {Vink}, J.~S., {et~al.}
  2014{\natexlab{b}}, \aap, 570, A38

\bibitem[{{Bj{\"o}rklund} {et~al.}(2022){Bj{\"o}rklund}, {Sundqvist}, {Singh},
  {Puls}, \& {Najarro}}]{Bjorklund2022}
{Bj{\"o}rklund}, R., {Sundqvist}, J.~O., {Singh}, S.~M., {Puls}, J., \&
  {Najarro}, F. 2022, arXiv e-prints, arXiv:2203.08218

\bibitem[{{Brands} {et~al.}(2022){Brands}, {de Koter}, {Bestenlehner},
  {Crowther}, {Sundqvist}, {Puls}, {Caballero-Nieves}, {Abdul-Masih},
  {Driessen}, {Garc{\'\i}a}, {Geen}, {Gr{\"a}fener}, {Hawcroft}, {Kaper},
  {Keszthelyi}, {Langer}, {Sana}, {Schneider}, {Shenar}, \&
  {Vink}}]{Brands2022}
{Brands}, S.~A., {de Koter}, A., {Bestenlehner}, J.~M., {et~al.} 2022, \aap,
  663, A36

\bibitem[{{Brinkman} {et~al.}(2023){Brinkman}, {Doherty}, {Pignatari}, {Pols},
  \& {Lugaro}}]{Brinkman2023}
{Brinkman}, H.~E., {Doherty}, C., {Pignatari}, M., {Pols}, O., \& {Lugaro}, M.
  2023, \apj, 951, 110

\bibitem[{{Bromm} {et~al.}(2002){Bromm}, {Coppi}, \& {Larson}}]{Bromm2002}
{Bromm}, V., {Coppi}, P.~S., \& {Larson}, R.~B. 2002, \apj, 564, 23

\bibitem[{{Cassinelli} {et~al.}(1981){Cassinelli}, {Mathis}, \&
  {Savage}}]{Cassinelli1981}
{Cassinelli}, J.~P., {Mathis}, J.~S., \& {Savage}, B.~D. 1981, Science, 212,
  1497

\bibitem[{{Clark} {et~al.}(2011){Clark}, {Glover}, {Klessen}, \&
  {Bromm}}]{Clark2011}
{Clark}, P.~C., {Glover}, S. C.~O., {Klessen}, R.~S., \& {Bromm}, V. 2011,
  \apj, 727, 110

\bibitem[{{Costa} {et~al.}(2021){Costa}, {Bressan}, {Mapelli}, {Marigo},
  {Iorio}, \& {Spera}}]{Costa2021}
{Costa}, G., {Bressan}, A., {Mapelli}, M., {et~al.} 2021, \mnras, 501, 4514

\bibitem[{{Crowther} {et~al.}(2010){Crowther}, {Schnurr}, {Hirschi}, {Yusof},
  {Parker}, {Goodwin}, \& {Kassim}}]{Crowther2010}
{Crowther}, P.~A., {Schnurr}, O., {Hirschi}, R., {et~al.} 2010, \mnras, 408,
  731

\bibitem[{{de Jager} {et~al.}(1988){de Jager}, {Nieuwenhuijzen}, \& {van der
  Hucht}}]{deJager1988}
{de Jager}, C., {Nieuwenhuijzen}, H., \& {van der Hucht}, K.~A. 1988, \aaps,
  72, 259

\bibitem[{{de Mink} {et~al.}(2009){de Mink}, {Pols}, {Langer}, \&
  {Izzard}}]{deMink2009}
{de Mink}, S.~E., {Pols}, O.~R., {Langer}, N., \& {Izzard}, R.~G. 2009, \aap,
  507, L1

\bibitem[{{Decressin} {et~al.}(2007){Decressin}, {Meynet}, {Charbonnel},
  {Prantzos}, \& {Ekstr{\"o}m}}]{Decressin2007}
{Decressin}, T., {Meynet}, G., {Charbonnel}, C., {Prantzos}, N., \&
  {Ekstr{\"o}m}, S. 2007, \aap, 464, 1029

\bibitem[{{Denissenkov} \& {Hartwick}(2014)}]{Denis2014}
{Denissenkov}, P.~A. \& {Hartwick}, F.~D.~A. 2014, \mnras, 437, L21

\bibitem[{{D'Ercole} {et~al.}(2010){D'Ercole}, {D'Antona}, {Ventura},
  {Vesperini}, \& {McMillan}}]{DErcole2010}
{D'Ercole}, A., {D'Antona}, F., {Ventura}, P., {Vesperini}, E., \& {McMillan},
  S. L.~W. 2010, \mnras, 407, 854

\bibitem[{{Dom{\'\i}nguez} {et~al.}(2022){Dom{\'\i}nguez}, {Pellegrini},
  {Klessen}, \& {Rahner}}]{Dominguez2022}
{Dom{\'\i}nguez}, R., {Pellegrini}, E.~W., {Klessen}, R.~S., \& {Rahner}, D.
  2022, arXiv e-prints, arXiv:2205.06209

\bibitem[{{Eggenberger} {et~al.}(2021){Eggenberger}, {Ekstr{\"o}m}, {Georgy},
  {Martinet}, {Pezzotti}, {Nandal}, {Meynet}, {Buldgen}, {Salmon},
  {Haemmerl{\'e}}, {Maeder}, {Hirschi}, {Yusof}, {Groh}, {Farrell}, {Murphy},
  \& {Choplin}}]{Eggenberger2021}
{Eggenberger}, P., {Ekstr{\"o}m}, S., {Georgy}, C., {et~al.} 2021, \aap, 652,
  A137

\bibitem[{{Eggenberger} {et~al.}(2008){Eggenberger}, {Meynet}, {Maeder},
  {Hirschi}, {Charbonnel}, {Talon}, \& {Ekstr{\"o}m}}]{Eggenberger2008}
{Eggenberger}, P., {Meynet}, G., {Maeder}, A., {et~al.} 2008, \apss, 316, 43

\bibitem[{{Eggenberger} {et~al.}(2010){Eggenberger}, {Miglio}, {Montalban},
  {Moreira}, {Noels}, {Meynet}, \& {Maeder}}]{Eggenberger2010}
{Eggenberger}, P., {Miglio}, A., {Montalban}, J., {et~al.} 2010, \aap, 509, A72

\bibitem[{{Ekstr{\"o}m} {et~al.}(2012){Ekstr{\"o}m}, {Georgy}, {Eggenberger},
  {Meynet}, {Mowlavi}, {Wyttenbach}, {Granada}, {Decressin}, {Hirschi},
  {Frischknecht}, {Charbonnel}, \& {Maeder}}]{Ekstrom2012}
{Ekstr{\"o}m}, S., {Georgy}, C., {Eggenberger}, P., {et~al.} 2012, \aap, 537,
  A146

\bibitem[{{Ekstr{\"o}m} {et~al.}(2008){Ekstr{\"o}m}, {Meynet}, {Maeder}, \&
  {Barblan}}]{Eks2008}
{Ekstr{\"o}m}, S., {Meynet}, G., {Maeder}, A., \& {Barblan}, F. 2008, \aap,
  478, 467

\bibitem[{{Farmer} {et~al.}(2019){Farmer}, {Renzo}, {de Mink}, {Marchant}, \&
  {Justham}}]{Farmer2019}
{Farmer}, R., {Renzo}, M., {de Mink}, S.~E., {Marchant}, P., \& {Justham}, S.
  2019, \apj, 887, 53

\bibitem[{{Farrell} {et~al.}(2022){Farrell}, {Groh}, {Meynet}, \&
  {Eldridge}}]{Farrell2022}
{Farrell}, E., {Groh}, J.~H., {Meynet}, G., \& {Eldridge}, J.~J. 2022, \mnras,
  512, 4116

\bibitem[{{Faucher-Gigu{\`e}re} {et~al.}(2008){Faucher-Gigu{\`e}re}, {Lidz},
  {Hernquist}, \& {Zaldarriaga}}]{FaucherGiguere2008}
{Faucher-Gigu{\`e}re}, C.-A., {Lidz}, A., {Hernquist}, L., \& {Zaldarriaga}, M.
  2008, \apj, 688, 85

\bibitem[{{Faucher-Gigu{\`e}re} {et~al.}(2009){Faucher-Gigu{\`e}re}, {Lidz},
  {Zaldarriaga}, \& {Hernquist}}]{FaucherGiguere2009}
{Faucher-Gigu{\`e}re}, C.-A., {Lidz}, A., {Zaldarriaga}, M., \& {Hernquist}, L.
  2009, \apj, 703, 1416

\bibitem[{{Figer}(2005)}]{Figer2005}
{Figer}, D.~F. 2005, \nat, 434, 192

\bibitem[{{Fragos} {et~al.}(2023){Fragos}, {Andrews}, {Bavera}, {Berry},
  {Coughlin}, {Dotter}, {Giri}, {Kalogera}, {Katsaggelos}, {Kovlakas},
  {Lalvani}, {Misra}, {Srivastava}, {Qin}, {Rocha}, {Rom{\'a}n-Garza}, {Serra},
  {Stahle}, {Sun}, {Teng}, {Trajcevski}, {Tran}, {Xing}, {Zapartas}, \&
  {Zevin}}]{Fragos2023}
{Fragos}, T., {Andrews}, J.~J., {Bavera}, S.~S., {et~al.} 2023, \apjs, 264, 45

\bibitem[{{Georgy} {et~al.}(2013){Georgy}, {Ekstr{\"o}m}, {Granada}, {Meynet},
  {Mowlavi}, {Eggenberger}, \& {Maeder}}]{Georgy2013}
{Georgy}, C., {Ekstr{\"o}m}, S., {Granada}, A., {et~al.} 2013, \aap, 553, A24

\bibitem[{{Georgy} {et~al.}(2012){Georgy}, {Ekstr{\"o}m}, {Meynet}, {Massey},
  {Levesque}, {Hirschi}, {Eggenberger}, \& {Maeder}}]{Georgy2012}
{Georgy}, C., {Ekstr{\"o}m}, S., {Meynet}, G., {et~al.} 2012, \aap, 542, A29

\bibitem[{{Georgy} {et~al.}(2014){Georgy}, {Saio}, \& {Meynet}}]{Georgy2014}
{Georgy}, C., {Saio}, H., \& {Meynet}, G. 2014, \mnras, 439, L6

\bibitem[{{Gieles} {et~al.}(2018){Gieles}, {Charbonnel}, {Krause},
  {H{\'e}nault-Brunet}, {Agertz}, {Lamers}, {Bastian}, {Gualandris}, {Zocchi},
  \& {Petts}}]{Gieles2018}
{Gieles}, M., {Charbonnel}, C., {Krause}, M. G.~H., {et~al.} 2018, \mnras, 478,
  2461

\bibitem[{{Gormaz-Matamala} {et~al.}(2023){Gormaz-Matamala}, {Cuadra},
  {Meynet}, \& {Cur{\'e}}}]{GormazMatamala2023}
{Gormaz-Matamala}, A.~C., {Cuadra}, J., {Meynet}, G., \& {Cur{\'e}}, M. 2023,
  \aap, 673, A109

\bibitem[{{Gr{\"a}fener} \& {Hamann}(2008)}]{Grafener2008}
{Gr{\"a}fener}, G. \& {Hamann}, W.~R. 2008, \aap, 482, 945

\bibitem[{{Gr{\"a}fener} {et~al.}(2012){Gr{\"a}fener}, {Owocki}, \&
  {Vink}}]{Graefener2012}
{Gr{\"a}fener}, G., {Owocki}, S.~P., \& {Vink}, J.~S. 2012, \aap, 538, A40

\bibitem[{{Greif} {et~al.}(2011){Greif}, {Springel}, {White}, {Glover},
  {Clark}, {Smith}, {Klessen}, \& {Bromm}}]{Greif2011}
{Greif}, T.~H., {Springel}, V., {White}, S. D.~M., {et~al.} 2011, \apj, 737, 75

\bibitem[{{Groenewegen}(2012{\natexlab{a}})}]{Groenewegen2012a}
{Groenewegen}, M.~A.~T. 2012{\natexlab{a}}, \aap, 540, A32

\bibitem[{{Groenewegen}(2012{\natexlab{b}})}]{Groenewegen2012b}
{Groenewegen}, M.~A.~T. 2012{\natexlab{b}}, \aap, 541, C3

\bibitem[{{Groh} {et~al.}(2019){Groh}, {Ekstr{\"o}m}, {Georgy}, {Meynet},
  {Choplin}, {Eggenberger}, {Hirschi}, {Maeder}, {Murphy}, {Boian}, \&
  {Farrell}}]{Groh2019}
{Groh}, J.~H., {Ekstr{\"o}m}, S., {Georgy}, C., {et~al.} 2019, \aap, 627, A24

\bibitem[{{Groh} {et~al.}(2014){Groh}, {Meynet}, {Ekstr{\"o}m}, \&
  {Georgy}}]{Groh2014}
{Groh}, J.~H., {Meynet}, G., {Ekstr{\"o}m}, S., \& {Georgy}, C. 2014, \aap,
  564, A30

\bibitem[{{Haehnelt} {et~al.}(2001){Haehnelt}, {Madau}, {Kudritzki}, \&
  {Haardt}}]{Haehnelt2001}
{Haehnelt}, M.~G., {Madau}, P., {Kudritzki}, R., \& {Haardt}, F. 2001, \apjl,
  549, L151

\bibitem[{{Higgins} {et~al.}(2023){Higgins}, {Vink}, {Hirschi}, {Laird}, \&
  {Sabhahit}}]{Higgins2023}
{Higgins}, E.~R., {Vink}, J.~S., {Hirschi}, R., {Laird}, A.~M., \& {Sabhahit},
  G.~N. 2023, arXiv e-prints, arXiv:2308.10941

\bibitem[{{Hirano} {et~al.}(2015){Hirano}, {Hosokawa}, {Yoshida}, {Omukai}, \&
  {Yorke}}]{Hirano2015}
{Hirano}, S., {Hosokawa}, T., {Yoshida}, N., {Omukai}, K., \& {Yorke}, H.~W.
  2015, \mnras, 448, 568

\bibitem[{{Hirano} {et~al.}(2014){Hirano}, {Hosokawa}, {Yoshida}, {Umeda},
  {Omukai}, {Chiaki}, \& {Yorke}}]{Hirano2014}
{Hirano}, S., {Hosokawa}, T., {Yoshida}, N., {et~al.} 2014, \apj, 781, 60

\bibitem[{{Hirschi}(2007)}]{Hirschi2007}
{Hirschi}, R. 2007, \aap, 461, 571

\bibitem[{{Ishii} {et~al.}(1999){Ishii}, {Ueno}, \& {Kato}}]{Ishii1999}
{Ishii}, M., {Ueno}, M., \& {Kato}, M. 1999, \pasj, 51, 417

\bibitem[{{Je{\v{r}}{\'a}bkov{\'a}} {et~al.}(2018){Je{\v{r}}{\'a}bkov{\'a}},
  {Hasani Zonoozi}, {Kroupa}, {Beccari}, {Yan}, {Vazdekis}, \&
  {Zhang}}]{Jerabkova2018}
{Je{\v{r}}{\'a}bkov{\'a}}, T., {Hasani Zonoozi}, A., {Kroupa}, P., {et~al.}
  2018, \aap, 620, A39

\bibitem[{Jiang {et~al.}(2023)Jiang, Chen, Tauris, Müller, \& Li}]{Jiang2023}
Jiang, L., Chen, W.-C., Tauris, T.~M., Müller, B., \& Li, X.-D. 2023, The
  Astrophysical Journal, 945, 90

\bibitem[{{Kaiser} {et~al.}(2020){Kaiser}, {Hirschi}, {Arnett}, {Georgy},
  {Scott}, \& {Cristini}}]{Kaiser2020}
{Kaiser}, E.~A., {Hirschi}, R., {Arnett}, W.~D., {et~al.} 2020, \mnras

\bibitem[{{Koen}(2006)}]{Koen2006}
{Koen}, C. 2006, \mnras, 365, 590

\bibitem[{{Krti{\v{c}}ka} {et~al.}(2011){Krti{\v{c}}ka}, {Owocki}, \&
  {Meynet}}]{Kri2011}
{Krti{\v{c}}ka}, J., {Owocki}, S.~P., \& {Meynet}, G. 2011, \aap, 527, A84

\bibitem[{{Maeder}(1997)}]{Maeder1997}
{Maeder}, A. 1997, \aap, 321, 134

\bibitem[{{Maeder} \& {Meynet}(2000)}]{Maeder2000}
{Maeder}, A. \& {Meynet}, G. 2000, \aap, 361, 159

\bibitem[{{Maeder} \& {Meynet}(2004)}]{Mader&Meynet2004TaylorSpruit}
{Maeder}, A. \& {Meynet}, G. 2004, \aap, 422, 225

\bibitem[{{Marchant} \& {Moriya}(2020)}]{Marchant2020}
{Marchant}, P. \& {Moriya}, T.~J. 2020, \aap, 640, L18

\bibitem[{{Marchant} {et~al.}(2019){Marchant}, {Renzo}, {Farmer}, {Pappas},
  {Taam}, {de Mink}, \& {Kalogera}}]{Marchant2019}
{Marchant}, P., {Renzo}, M., {Farmer}, R., {et~al.} 2019, \apj, 882, 36

\bibitem[{{Martinet} {et~al.}(2021){Martinet}, {Meynet}, {Ekstr{\"o}m},
  {Sim{\'o}n-D{\'\i}az}, {Holgado}, {Castro}, {Georgy}, {Eggenberger},
  {Buldgen}, {Salmon}, {Hirschi}, {Groh}, {Farrell}, \&
  {Murphy}}]{Martinet2021}
{Martinet}, S., {Meynet}, G., {Ekstr{\"o}m}, S., {et~al.} 2021, \aap, 648, A126

\bibitem[{{Martinet} {et~al.}(2022){Martinet}, {Meynet}, {Nandal},
  {Ekstr{\"o}m}, {Georgy}, {Haemmerl{\'e}}, {Hirschi}, {Yusof}, {Gounelle}, \&
  {Dwarkadas}}]{Martinet2022}
{Martinet}, S., {Meynet}, G., {Nandal}, D., {et~al.} 2022, \aap, 664, A181

\bibitem[{{Massey} {et~al.}(2023){Massey}, {Neugent}, {Ekstr{\"o}m}, {Georgy},
  \& {Meynet}}]{Massey2023}
{Massey}, P., {Neugent}, K.~F., {Ekstr{\"o}m}, S., {Georgy}, C., \& {Meynet},
  G. 2023, \apj, 942, 69

\bibitem[{{McEvoy} {et~al.}(2015){McEvoy}, {Dufton}, {Evans}, {Kalari},
  {Markova}, {Sim{\'o}n-D{\'\i}az}, {Vink}, {Walborn}, {Crowther}, {de Koter},
  {de Mink}, {Dunstall}, {H{\'e}nault-Brunet}, {Herrero}, {Langer}, {Lennon},
  {Ma{\'\i}z Apell{\'a}niz}, {Najarro}, {Puls}, {Sana}, {Schneider}, \&
  {Taylor}}]{McEvoy2015}
{McEvoy}, C.~M., {Dufton}, P.~L., {Evans}, C.~J., {et~al.} 2015, \aap, 575, A70

\bibitem[{{Murphy} {et~al.}(2021){Murphy}, {Groh}, {Farrell}, {Meynet},
  {Ekstr{\"o}m}, {Tsiatsiou}, {Hackett}, \& {Martinet}}]{Murphy2021}
{Murphy}, L.~J., {Groh}, J.~H., {Farrell}, E., {et~al.} 2021, \mnras

\bibitem[{{Nugis} \& {Lamers}(2000)}]{Nugis2000}
{Nugis}, T. \& {Lamers}, H.~J.~G.~L.~M. 2000, \aap, 360, 227

\bibitem[{{Oey} \& {Clarke}(2005)}]{Oey2005}
{Oey}, M.~S. \& {Clarke}, C.~J. 2005, \apjl, 620, L43

\bibitem[{{Ram{\'\i}rez-Agudelo} {et~al.}(2017){Ram{\'\i}rez-Agudelo}, {Sana},
  {de Koter}, {Tramper}, {Grin}, {Schneider}, {Langer}, {Puls}, {Markova},
  {Bestenlehner}, {Castro}, {Crowther}, {Evans}, {Garc{\'\i}a}, {Gr{\"a}fener},
  {Herrero}, {van Kempen}, {Lennon}, {Ma{\'\i}z Apell{\'a}niz}, {Najarro},
  {Sab{\'\i}n-Sanjuli{\'a}n}, {Sim{\'o}n-D{\'\i}az}, {Taylor}, \&
  {Vink}}]{RamirezAgudelo2017}
{Ram{\'\i}rez-Agudelo}, O.~H., {Sana}, H., {de Koter}, A., {et~al.} 2017, \aap,
  600, A81

\bibitem[{{Rivinius} {et~al.}(2013){Rivinius}, {Carciofi}, \&
  {Martayan}}]{Rivinius2013}
{Rivinius}, T., {Carciofi}, A.~C., \& {Martayan}, C. 2013, \aapr, 21, 69

\bibitem[{{Sabhahit} {et~al.}(2022){Sabhahit}, {Vink}, {Higgins}, \&
  {Sander}}]{Sabhahit2022}
{Sabhahit}, G.~N., {Vink}, J.~S., {Higgins}, E.~R., \& {Sander}, A. A.~C. 2022,
  \mnras, 514, 3736

\bibitem[{{Sabhahit} {et~al.}(2023){Sabhahit}, {Vink}, {Sander}, \&
  {Higgins}}]{Sabhahit2023}
{Sabhahit}, G.~N., {Vink}, J.~S., {Sander}, A. A.~C., \& {Higgins}, E.~R. 2023,
  \mnras, 524, 1529

\bibitem[{{Sab{\'\i}n-Sanjuli{\'a}n} {et~al.}(2017){Sab{\'\i}n-Sanjuli{\'a}n},
  {Sim{\'o}n-D{\'\i}az}, {Herrero}, {Puls}, {Schneider}, {Evans}, {Garcia},
  {Najarro}, {Brott}, {Castro}, {Crowther}, {de Koter}, {de Mink},
  {Gr{\"a}fener}, {Grin}, {Holgado}, {Langer}, {Lennon}, {Ma{\'\i}z
  Apell{\'a}niz}, {Ram{\'\i}rez-Agudelo}, {Sana}, {Taylor}, {Vink}, \&
  {Walborn}}]{SabinSanjulian2017}
{Sab{\'\i}n-Sanjuli{\'a}n}, C., {Sim{\'o}n-D{\'\i}az}, S., {Herrero}, A.,
  {et~al.} 2017, \aap, 601, A79

\bibitem[{{Sab{\'\i}n-Sanjuli{\'a}n} {et~al.}(2014){Sab{\'\i}n-Sanjuli{\'a}n},
  {Sim{\'o}n-D{\'\i}az}, {Herrero}, {Walborn}, {Puls}, {Ma{\'\i}z
  Apell{\'a}niz}, {Evans}, {Brott}, {de Koter}, {Garcia}, {Markova}, {Najarro},
  {Ram{\'\i}rez-Agudelo}, {Sana}, {Taylor}, \& {Vink}}]{SabinSanjulian2014}
{Sab{\'\i}n-Sanjuli{\'a}n}, C., {Sim{\'o}n-D{\'\i}az}, S., {Herrero}, A.,
  {et~al.} 2014, \aap, 564, A39

\bibitem[{{Sander} \& {Vink}(2020)}]{Sander2020}
{Sander}, A. A.~C. \& {Vink}, J.~S. 2020, \mnras, 499, 873

\bibitem[{{Sanyal} {et~al.}(2015){Sanyal}, {Grassitelli}, {Langer}, \&
  {Bestenlehner}}]{Sanyal2015}
{Sanyal}, D., {Grassitelli}, L., {Langer}, N., \& {Bestenlehner}, J.~M. 2015,
  \aap, 580, A20

\bibitem[{{Schaller} {et~al.}(1992){Schaller}, {Schaerer}, {Meynet}, \&
  {Maeder}}]{Schaller1992}
{Schaller}, G., {Schaerer}, D., {Meynet}, G., \& {Maeder}, A. 1992, \aaps, 96,
  269

\bibitem[{{Schneider} {et~al.}(2018){Schneider}, {Ram{\'\i}rez-Agudelo},
  {Tramper}, {Bestenlehner}, {Castro}, {Sana}, {Evans},
  {Sab{\'\i}n-Sanjuli{\'a}n}, {Sim{\'o}n-D{\'\i}az}, {Langer}, {Fossati},
  {Gr{\"a}fener}, {Crowther}, {de Mink}, {de Koter}, {Gieles}, {Herrero},
  {Izzard}, {Kalari}, {Klessen}, {Lennon}, {Mahy}, {Ma{\'\i}z Apell{\'a}niz},
  {Markova}, {van Loon}, {Vink}, \& {Walborn}}]{Schneider2018}
{Schneider}, F.~R.~N., {Ram{\'\i}rez-Agudelo}, O.~H., {Tramper}, F., {et~al.}
  2018, \aap, 618, A73

\bibitem[{{Scott} {et~al.}(2021){Scott}, {Hirschi}, {Georgy}, {Arnett},
  {Meakin}, {Kaiser}, {Ekstr{\"o}m}, \& {Yusof}}]{Scott2021}
{Scott}, L.~J.~A., {Hirschi}, R., {Georgy}, C., {et~al.} 2021, \mnras, 503,
  4208

\bibitem[{{Sibony} {et~al.}(2022){Sibony}, {Liu}, {Simmonds}, {Meynet}, \&
  {Bromm}}]{Sibony2022}
{Sibony}, Y., {Liu}, B., {Simmonds}, C., {Meynet}, G., \& {Bromm}, V. 2022,
  \aap, 666, A199

\bibitem[{{Simaz Bunzel} {et~al.}(2023){Simaz Bunzel}, {Garc{\'\i}a}, {Combi},
  \& {Chaty}}]{SimazBunzel2023}
{Simaz Bunzel}, A., {Garc{\'\i}a}, F., {Combi}, J.~A., \& {Chaty}, S. 2023,
  \aap, 670, A80

\bibitem[{{Sim{\'o}n-D{\'\i}az} {et~al.}(2017){Sim{\'o}n-D{\'\i}az}, {Godart},
  {Castro}, {Herrero}, {Aerts}, {Puls}, {Telting}, \&
  {Grassitelli}}]{SimonDiaz2017}
{Sim{\'o}n-D{\'\i}az}, S., {Godart}, M., {Castro}, N., {et~al.} 2017, \aap,
  597, A22

\bibitem[{{Song} \& {Liu}(2023)}]{Song2023}
{Song}, C.-Y. \& {Liu}, T. 2023, \apj, 952, 156

\bibitem[{{Spruit}(2002)}]{Spruit2002}
{Spruit}, H.~C. 2002, \aap, 381, 923

\bibitem[{{Stacy} \& {Bromm}(2013)}]{Stacy2013}
{Stacy}, A. \& {Bromm}, V. 2013, \mnras, 433, 1094

\bibitem[{{Stacy} {et~al.}(2016){Stacy}, {Bromm}, \& {Lee}}]{Stacy2016}
{Stacy}, A., {Bromm}, V., \& {Lee}, A.~T. 2016, \mnras, 462, 1307

\bibitem[{{Stacy} {et~al.}(2010){Stacy}, {Greif}, \& {Bromm}}]{Stacy2010}
{Stacy}, A., {Greif}, T.~H., \& {Bromm}, V. 2010, \mnras, 403, 45

\bibitem[{{Stothers}(1999)}]{Stothers1999}
{Stothers}, R.~B. 1999, \mnras, 305, 365

\bibitem[{{Susa} {et~al.}(2014){Susa}, {Hasegawa}, \& {Tominaga}}]{Susa2014}
{Susa}, H., {Hasegawa}, K., \& {Tominaga}, N. 2014, \apj, 792, 32

\bibitem[{{Sz{\'e}csi} {et~al.}(2018){Sz{\'e}csi}, {Mackey}, \&
  {Langer}}]{Szecsi2018}
{Sz{\'e}csi}, D., {Mackey}, J., \& {Langer}, N. 2018, \aap, 612, A55

\bibitem[{{Timmes} \& {Swesty}(2000)}]{Timmes2000}
{Timmes}, F.~X. \& {Swesty}, F.~D. 2000, \apjs, 126, 501

\bibitem[{{Turk} {et~al.}(2009){Turk}, {Abel}, \& {O'Shea}}]{Turk2009}
{Turk}, M.~J., {Abel}, T., \& {O'Shea}, B. 2009, Science, 325, 601

\bibitem[{{van Loon} {et~al.}(2005){van Loon}, {Cioni}, {Zijlstra}, \&
  {Loup}}]{vanLoon2005}
{van Loon}, J.~T., {Cioni}, M. R.~L., {Zijlstra}, A.~A., \& {Loup}, C. 2005,
  \aap, 438, 273

\bibitem[{{Vanbeveren} {et~al.}(2012){Vanbeveren}, {Mennekens}, \& {De
  Greve}}]{Vanbeveren2012}
{Vanbeveren}, D., {Mennekens}, N., \& {De Greve}, J.~P. 2012, \aap, 543, A4

\bibitem[{{Vink}(2018)}]{Vink2018}
{Vink}, J.~S. 2018, \aap, 615, A119

\bibitem[{{Vink}(2021)}]{Vink2021}
{Vink}, J.~S. 2021, arXiv e-prints, arXiv:2109.08164

\bibitem[{{Vink} {et~al.}(2001){Vink}, {de Koter}, \& {Lamers}}]{Vink2001}
{Vink}, J.~S., {de Koter}, A., \& {Lamers}, H.~J.~G.~L.~M. 2001, \aap, 369, 574

\bibitem[{{Weidner} \& {Kroupa}(2004)}]{Weidner2004}
{Weidner}, C. \& {Kroupa}, P. 2004, \mnras, 348, 187

\bibitem[{{Wise} {et~al.}(2014){Wise}, {Demchenko}, {Halicek}, {Norman},
  {Turk}, {Abel}, \& {Smith}}]{Wise2014}
{Wise}, J.~H., {Demchenko}, V.~G., {Halicek}, M.~T., {et~al.} 2014, \mnras,
  442, 2560

\bibitem[{{Wollenberg} {et~al.}(2020){Wollenberg}, {Glover}, {Clark}, \&
  {Klessen}}]{Wollenberg2020}
{Wollenberg}, K. M.~J., {Glover}, S. C.~O., {Clark}, P.~C., \& {Klessen}, R.~S.
  2020, \mnras, 494, 1871

\bibitem[{{Woosley} \& {Heger}(2021)}]{Woosley2021}
{Woosley}, S.~E. \& {Heger}, A. 2021, \apjl, 912, L31

\bibitem[{{Yang} {et~al.}(2023){Yang}, {Bonanos}, {Jiang}, {Zapartas}, {Gao},
  {Ren}, {Lam}, {Wang}, {Maravelias}, {Gavras}, {Wang}, {Chen}, {Tramper}, {de
  Wit}, {Chen}, {Wen}, {Liu}, {Tian}, {Antoniadis}, \& {Luo}}]{Yang2023}
{Yang}, M., {Bonanos}, A.~Z., {Jiang}, B., {et~al.} 2023, \aap, 676, A84

\bibitem[{{Yusof} {et~al.}(2022){Yusof}, {Hirschi}, {Eggenberger},
  {Ekstr{\"o}m}, {Georgy}, {Sibony}, {Crowther}, {Meynet}, {Kassim}, {Harun},
  {Maeder}, {Groh}, {Farrell}, \& {Murphy}}]{Yusof2022}
{Yusof}, N., {Hirschi}, R., {Eggenberger}, P., {et~al.} 2022, \mnras, 511, 2814

\bibitem[{{Yusof} {et~al.}(2013){Yusof}, {Hirschi}, {Meynet}, {Crowther},
  {Ekstr{\"o}m}, {Frischknecht}, {Georgy}, {Abu Kassim}, \&
  {Schnurr}}]{Yusof2013}
{Yusof}, N., {Hirschi}, R., {Meynet}, G., {et~al.} 2013, \mnras, 433, 1114

\bibitem[{{Zahn}(1992)}]{Zahn1992}
{Zahn}, J.~P. 1992, \aap, 265, 115

\end{thebibliography}

\end{document}